# Collaborative planning of integrated hydrogen energy chain multi-energy systems: A review

*Xinning Yi*[a], *Tianguang Lu*[a,b]*, *Jing Li*[c], *Shaocong Wu*[a]

[a]School of Electrical Engineering, Shandong University, Jinan 250061, China
[b]School of Engineering and Applied Sciences and Harvard China Project, Harvard University, Cambridge, MA 02138, United States
[c]State Grid Coporation of China, Beijing 100032, China
*tlu@sdu.edu.cn

---

**Abstract:** Most planning of the traditional hydrogen energy supply chain (HSC) focuses on the storage and transportation links between production and consumption ends. It ignores the energy flows and interactions between each link, making it unsuitable for energy system planning analysis. Therefore, we propose the concept of a hydrogen energy chain (HEC) based on the HSC, which emphasizes the interactions between different types of energy flows in the production, compression, storage, transportation, and application links of hydrogen. The HEC plays a crucial role in mitigating fluctuations of renewable energy and facilitating the optimal allocation of heterogeneous energy sources across time and space. Effective collaborative planning models that consider HEC are essential for the optimal configuration of multi-energy systems (MESs), which guarantees high-efficiency operation and the economic and environmental friendliness of the system. This paper presents a systematic review of recent articles on collaborative planning of integrated hydrogen energy chain multi-energy systems (HEC-MESs). First, we introduce the basic framework of HEC-MES, focusing on the current research status of the production, compression, storage, transportation, and application links in HEC. Furthermore, we review technology types of hydrogen energy for planning and summarize the typical forms of HEC in MESs. Then, the following sections outline the models and methods for collaborative planning of HEC-MES. They include detailed analyses of covered sector types, spatial and temporal scopes of planning, uncertainties, model formulations, and solution methods. Finally, the paper concludes by summarizing the research gaps identified in current articles and outlining directions for future research.

---

* *Corresponding author*
E-mail address: tlu@sdu.edu.cn (T. Lu).

Nomenclature

| | | | |
|---|---|---|---|
| HSC | Hydrogen energy supply chain | DRO | Distributed robust optimization |
| HEC | Hydrogen energy chain | SMR | Steam methane reforming |
| HEC-MES | Integrated HEC multi-energy systems | MILP | Mixed integer linear programming |
| WT | Wind turbine | SC | Salt cavern |
| PV | Photovoltaic panel | MC | Monte Carlo |
| MES | Multi-energy system | H2P | Hydrogen-to-power |
| EC | Electrolysis cell | SOFC | Solid oxide fuel cell |
| IEA | International Energy Agency | PEMFC | Proton exchange membrane fuel cell |
| HS | Hydrogen energy storage | HV | Hydrogen vehicle |
| HT | Hydrogen-fired turbine | SO | Stochastic optimization |
| FC | Fuel cell | RO | Robust optimization |
| HFCV | Hydrogen fuel cell vehicle | TANPC | Total annual net present value cost |
| IES | Integrated energy system | MINLP | Mixed integer non-linear programming |
| CHP | Combined heat and power | LP | Linear programming |
| CCS | Carbon capture and sequestration | PSO | Particle swarm optimization |
| HST | Hydrogen storage tank | ERS | Elite retention strategy |
| HRS | Hydrogen refueling station | SDT | Strong duality theory |
| HB | Hydrogen-fired boiler | NNB | Newton's numerical barrier method |
| GT | Gas turbine | MIQCP | Mixed integer quadratic programming |

# 1. Introduction

Today, fossil fuels remain dominant in the global energy supply, accounting for approximately 82% of global primary energy consumption in 2023 [1]. However, the massive consumption and non-renewable nature of fossil fuels are leading to their depletion. In addition, carbon dioxide emissions and pollutants (e.g., nitrogen and sulfur) emitted from combustion cause irrevocable and destructive damage to the environment [2]. Thus, there is an urgent need to transition to a low-carbon and sustainable global energy future. The Paris Agreement commits to limiting the global temperature rise to 2°C and emphasizes that the world will peak greenhouse gas emissions as soon as possible [3]. The vigorous development of renewable energy has become an important energy strategy for many countries. According to the International Energy Agency (IEA) [4], the installed capacity of wind turbines (WTs) and photovoltaics (PVs) has increased by 9.9% and 23.8%, respectively, in 2022 compared to that in 2023. The total global installed capacity of WTs and PVs is expected to reach 16.0 TW and 5.9 TW, respectively in 2050.

* *Corresponding author.*
E-mail address: tlu@sdu.edu.cn (T. Lu).

With the integration of large-scale renewable energy, increasing stochasticity and volatility pose significant challenges to the safety and economy of energy system operation [5]. Nowadays, the energy system is characterized by an imbalance between the geographical distribution and development of renewable energy resources, a seasonal mismatch between energy supply and demand, and a lack of flexibility [6-8]. The construction of multi-energy systems (MESs) has become an effective approach for addressing the above problems and challenges. Compared with traditional energy systems, MESs can realize synergistic multi-energy supply and gradient utilization. Furthermore, it can improve overall flexibility and efficiency while meeting energy demands [9].

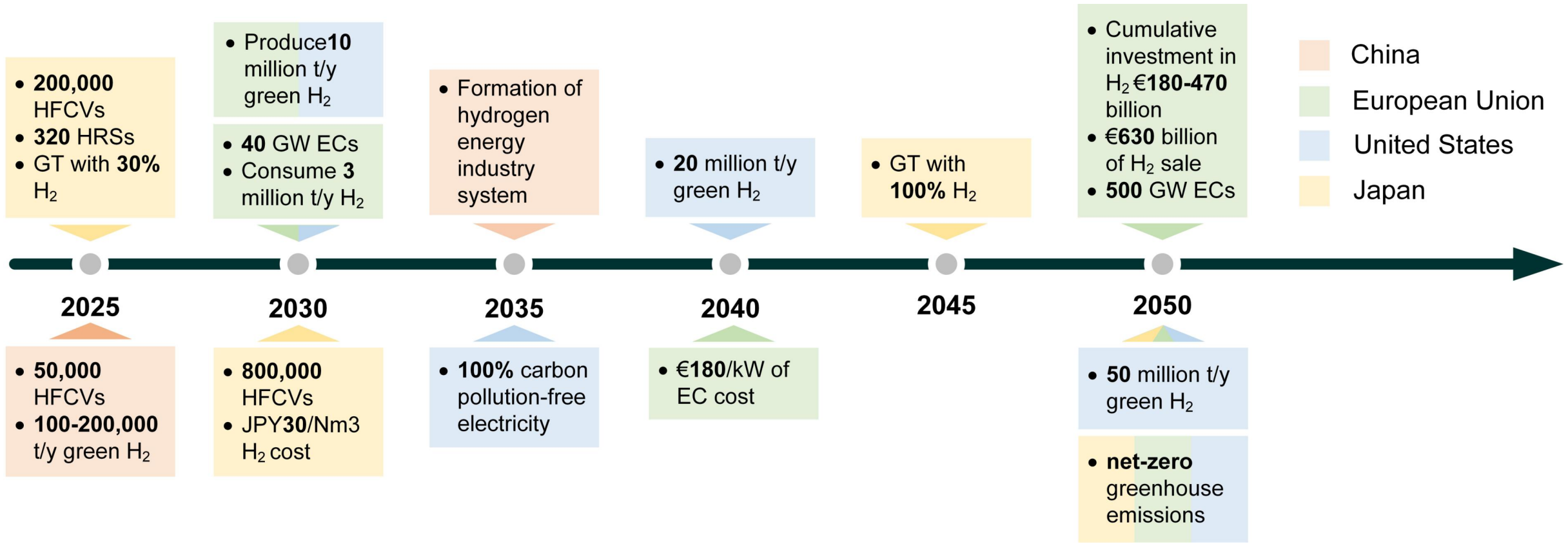

**Fig. 1.** Representative strategic routes for hydrogen energy development

Hydrogen energy is considered as a significant carrier for the deep decarbonization of energy systems in the future due to its high combustion value, low carbon emissions, and high flexibility of storage/conversion [10]. Electrolysis cells (ECs) can convert surplus wind and solar energy into hydrogen [11]. According to the IEA forecast, the global electricity used for hydrogen production will exceed 8,000 TWh in 2050, accounting for more than 60% of the energy source for hydrogen production [4]. Hydrogen energy storage (HS) is a large-scale and long-term form of energy storage [12]. When combined with hydrogen transportation, it allows for the optimal allocation of energy across geographical and seasonal boundaries. Hydrogen energy can also generate clean electricity and heat using equipment such as hydrogen-fired turbines (HTs) and fuel cells (FCs) [13-14]. It can be concluded that hydrogen will play a crucial role in various future application scenarios. With this background, governments worldwide have successively formulated strategic routes for hydrogen energy development and actively promoted hydrogen energy technology research and industrial deployment, among which China [15], the European Union [16], the United States [17], and Japan [18] are representative. Figure 1 illustrates representative strategic routes for hydrogen energy development. The global consensus for the development of hydrogen energy is to promote the construction of a complete top-down hydrogen energy industry chain.

The integration of hydrogen energy contributes to the balance of heterogeneous energy supply and demand, as well as the flexibility enhancement within MESs [19]. However, there are challenges due to the complex relationship between energy coupling [20] and multiple spatiotemporal scale characteristics [21]. This leads to difficulties in the coordinated operation of different devices and synergistic mutualization of energy flows in MES planning. Traditional planning methods cannot meet the planning needs of multi-energy flow and diversified scenarios, which easily causes planning deviation, high investment costs, and energy resource waste. Collaborative planning between different energy networks and equipment is an effective measure for addressing the above challenges. It can quantitatively depict the differences in the characteristics of different energy flows, accurately describe the dynamics and interactions of equipment, and effectively improve the reliability and economy of the planning results. Recently, researchers have conducted numerous studies to explore collaborative planning methods to enhance the economy, performance, and feasibility of MESs containing hydrogen. Literature [22] studied zero-carbon emission energy systems in Europe and showed that collaborative planning considering ECs and HSs can reduce the hydrogen cost by 10 €/MWh. Peng Hou et al. [23] proposed a coupled system combining offshore wind power and hydrogen energy. The authors noted that the utilization of surplus wind energy for hydrogen production through ECs is the optimal investment strategy for maximizing the profit of wind farms. Research [24] established a hydrogen storage system model considering electric-heat coupling and conducted a collaborative planning analysis of an integrated energy system (IES). This verified that electric-hydrogen-heat conversion contributes to improving energy utilization efficiency.

To date, many researchers have conducted overviews of a specific type of energy system containing hydrogen (e.g., electricity-gas-hydrogen IESs) and related planning techniques. Reference [25] provided a comprehensive review of hydrogen-based combined heat and power (CHP) systems, covering equipment technology, hydrogen management, and system design. This review suggested that rational spatial planning can reduce system costs and energy losses. A.Z. Arsad et al. [26] provided a detailed review of the issues, difficulties, and challenges of IESs with HSs based on a comprehensive analysis of highly cited papers. Literature [27] summarized the modeling and scope of the electricity-gas-hydrogen energy system planning model based on recent related research and provided an in-depth analysis of the system characteristics and technical features. However, the above reviews analyzed only a single type of energy system and did not investigate complete hydrogen links with diverse energy conversion relationships. Furthermore, study [26] and [27] focused more on system-wide overviews without analyzing planning in sufficient depth. Currently, some papers offer an extensive review of hydrogen energy supply chain (HSC) design and planning. Reference [28] provided a comprehensive overview of the fundamental concepts, development routes, and construction scenarios of HSCs. It also analyzed the planning models and methods of related infrastructures in a categorized manner. In addition, Fabio Sgarbossa et al. [29] spotlighted the HSCs for hydrogen production from renewable power generation and proposed a planning matrix for renewable HSC planning based on a systematic analysis of existing studies. Literature [30] logically reviewed the research on HSC network design models, provided in-depth analyses and classified summaries around modeling and solution strategies, and pointed out the shortcomings of the existing research and the possible problems of future research. However, the above reviews focused on the planning of HSCs, especially on the storage and transportation link between production and consumption ends, which is weakly related to the energy system. They paid little attention to the energy flows and interactions between the various links. Therefore, integrating the HSC with the energy system for collaborative planning is not recommended. Furthermore, several papers have provided a comprehensive overview of only specific hydrogen energy technologies (e.g., ECs [31], HSs [32], and FCs [33]) and hydrogen energy systems [34].

To fill the above gap, the hydrogen energy chain (HEC) is proposed based on HSC and focuses on the interactions between energy flows in the production, compression, storage, transportation, and application links of hydrogen. This paper comprehensively reviews recent research on the collaborative planning of integrated hydrogen energy chain multi-energy systems (HEC-MESs), emphasizing the close relationship between HEC and the energy sector. This study aims to analyze the collaborative planning of the HEC with other energy networks and equipment in MESs and to explore the role of coupled hydrogen energy conversion. First, this review explores the basic concepts and typical forms of the HEC integrated into the MES and provides a systematic review of its top-down production, compression, storage, transport, and application aspects. The survey of equipment and its features related to the HECs provides a more comprehensive understanding. Next, it summarizes the general forms of the existing research on collaborative planning models, including their covered sector types (e.g., the electric power and traffic sectors), spatial and temporal scopes of planning, and uncertainties. Additionally, this paper provides an in-depth analysis of the objective function, the forms of the formulation, and its characteristics, as well as a detailed classification and analysis of the model's solution technologies. Finally, the paper identifies research gaps and suggests future directions. Overall, this article presents a comprehensive review of the research on collaborative planning of HEC-MESs.

This study provides inspiration for researchers in the field and offers a reference for future low-carbon energy system planning scenarios.

The remainder of this paper is divided into two main parts (Part 1 includes Sections 2 and 3, and Part 2 includes Sections 4 and 5), as shown in Fig. 2. It is organized as follows: 1. The basic configuration of HEC-MES includes the HEC and the composition of its various links (Section 2), as well as the types of technologies and basic models of the equipment related to the HEC in MES (Section 3). 2. Collaborative planning for HEC-MES, including the basic features (Section 4) and solution technologies (Section 5). In Section 6, the paper summarizes the contents of this paper, discusses the limitations of existing research and discusses future guideline directions.

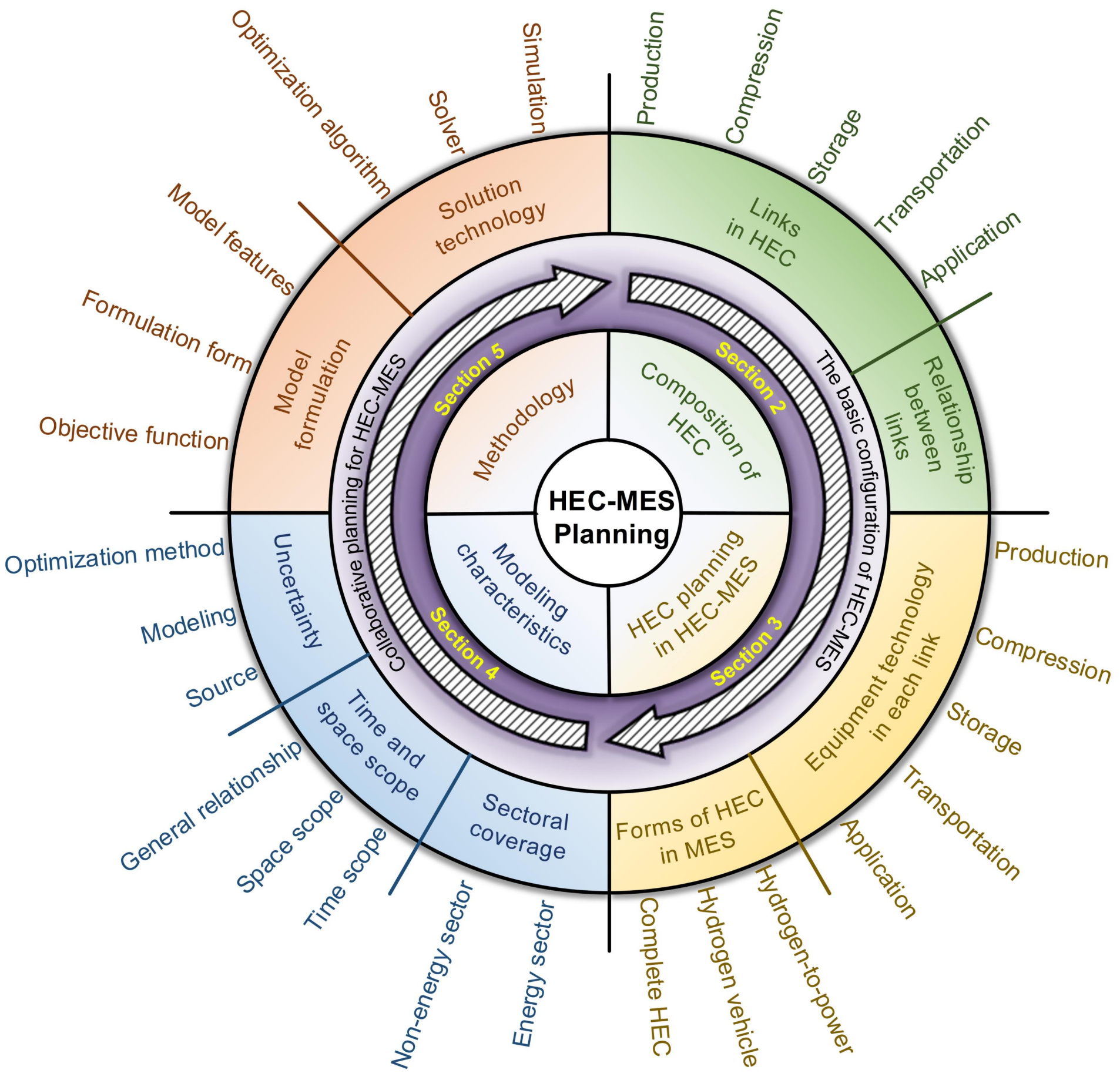

**Fig. 2.** The basic content structure of this review

# 2. Composition of HEC

HEC typically involves production, compression, storage, transportation, and application, each in a different form. This section focuses on the current development of each link in HEC, the type of technology, and its role in the energy sector. Then, we explain the concept of a complete HEC in a comprehensive top-down manner and analyze the connections between the links.

## *2.1 Hydrogen production link*

Hydrogen production is the primary link in HEC and is also a prerequisite for the future large-scale utilization of hydrogen energy. Figure 3 illustrates the technology categorization of hydrogen production. According to the IEA [35], global hydrogen production in 2022 has increased by 3% compared to that in 2021, reaching 95 million tons. However, less than 1% of this is clean hydrogen production. Currently, the majority of hydrogen production relies on fossil energy reforming, with coal gasification and natural gas reforming accounting for 62% and 21%, respectively. Additionally, 16% of hydrogen production comes from industrial by-products such as coke oven gas and chlor-alkali tail gas [10]. The World Energy Council [36] categorizes the produced hydrogen based on its carbon emissions, using the colors grey, blue, and green to represent descending levels of emissions. Grey and blue hydrogen are usually distinguished by whether the production link utilizes carbon capture and sequestration (CCS) technology [37]. Green hydrogen refers to hydrogen produced with zero-carbon emissions. It is essential to conduct research on large-scale clean hydrogen production technology to achieve decarbonization. Generally, clean hydrogen production technologies can be categorized into direct production from renewable energy and clean electricity electrolysis of water. The former technologies are in the experimental development stage and are not yet suitable for large-scale applications, including biomass, solar photocatalytic, pyrolysis hydrogen production, and other technologies [38]. In contrast, the latter technology is relatively mature on an industrial scale, but there is a significant cost gap compared to traditional hydrogen production methods. The technology widely used for hydrogen production in energy system research is EC using wind and solar generation [39]. This technology can meet the demand for green hydrogen while smoothing out renewable energy fluctuations.

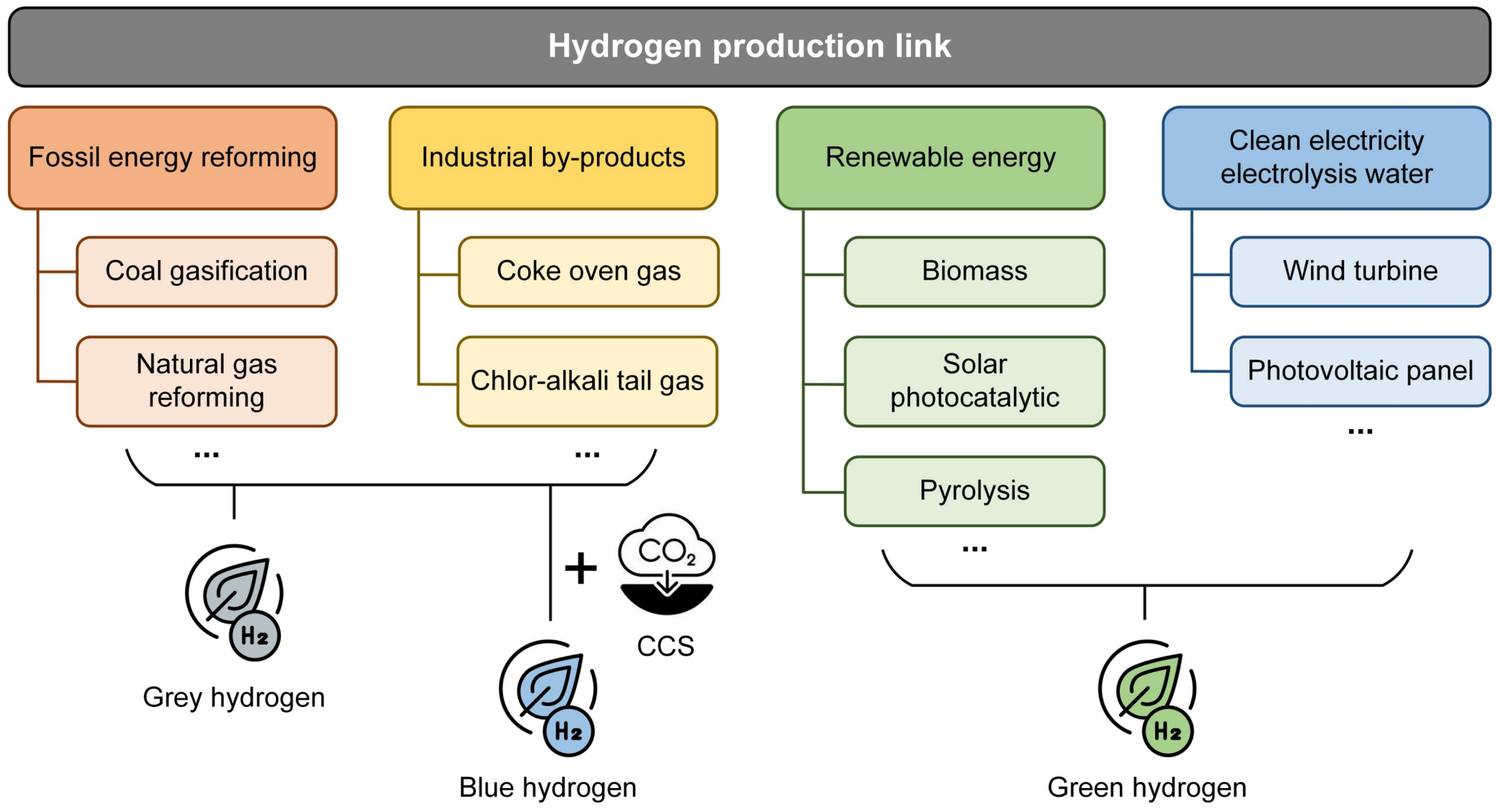

**Fig. 3.** Classification of technologies in hydrogen production link

## *2.2 Hydrogen compression link*

At normal temperature and pressure, hydrogen is a gas with a low volumetric energy density. Therefore, it is generally compressed using various compression techniques to the appropriate state for storage, transportation, and application. Compression is closely linked to the storage and transportation links, and their basic technical classifications and forms of hydrogen energy flow are presented in Fig. 4. Hydrogen compression technology can be classified into mechanical and non-mechanical compression methods based on the types of compressors used. Mechanical compression is commonly used in current energy systems to pressurize low-pressure hydrogen into a high-pressure gaseous state [40]. It has the advantages of low cost, low energy consumption, and mature technology. Non-mechanical compression methods include cryogenic liquefaction, metal hydride, and electrochemical compression. Compared with mechanical compression, non-mechanical compression offers greater compression efficiency and safety [41]. However, considering technical challenges and associated costs, the large-scale implementation of non-mechanical compression in energy systems is not yet feasible. To accurately characterize the role of hydrogen in energy systems, it is essential to consider hydrogen losses and compressor energy consumption during compression.

## *2.3 Hydrogen storage link*

The hydrogen storage link is an essential component of the HEC and serves as the basis for long-term energy storage and large-scale hydrogen transfer. Hydrogen can be stored in three material forms: gaseous, liquid, and solid. According to the theory, hydrogen storage technology can be divided into physical and chemical types. The former includes high-pressure gaseous, low-temperature liquid, and activated carbon adsorption hydrogen storage, while the latter involves metal hydrides and organic liquid hydride storage [42-43]. While low-temperature liquid hydrogen storage has the advantage of high energy density, ensuring its safety in terms of adiabatic requirements is difficult and expensive. Furthermore, technical challenges such as excessive energy consumption during the liquefaction process still require resolution [44]. Solid hydrogen storage has a higher energy density than liquid hydrogen storage and has the advantages of high safety, low storage pressure, and high purity. Therefore, it is widely regarded as the best method for hydrogen storage [45]. However, solid hydrogen storage is not yet available for large-scale applications because of technical difficulties and expensive cost constraints. Currently, the most widespread method applied in energy systems is high-pressure gaseous hydrogen storage. This method offers advantages such as fast charging/discharging speed, simple structure, and mature technology [46]. To improve the operational flexibility of the energy system, mobile hydrogen storage has been proposed since fixed hydrogen storage cannot transfer energy on a spatial scale [47]. Compared with traditional electrochemical energy storage, the long-term storage characteristics of hydrogen can effectively address the contradiction of the seasonal imbalance between renewable energy output and load demand [7]. It enables the optimal allocation of energy resources over time.

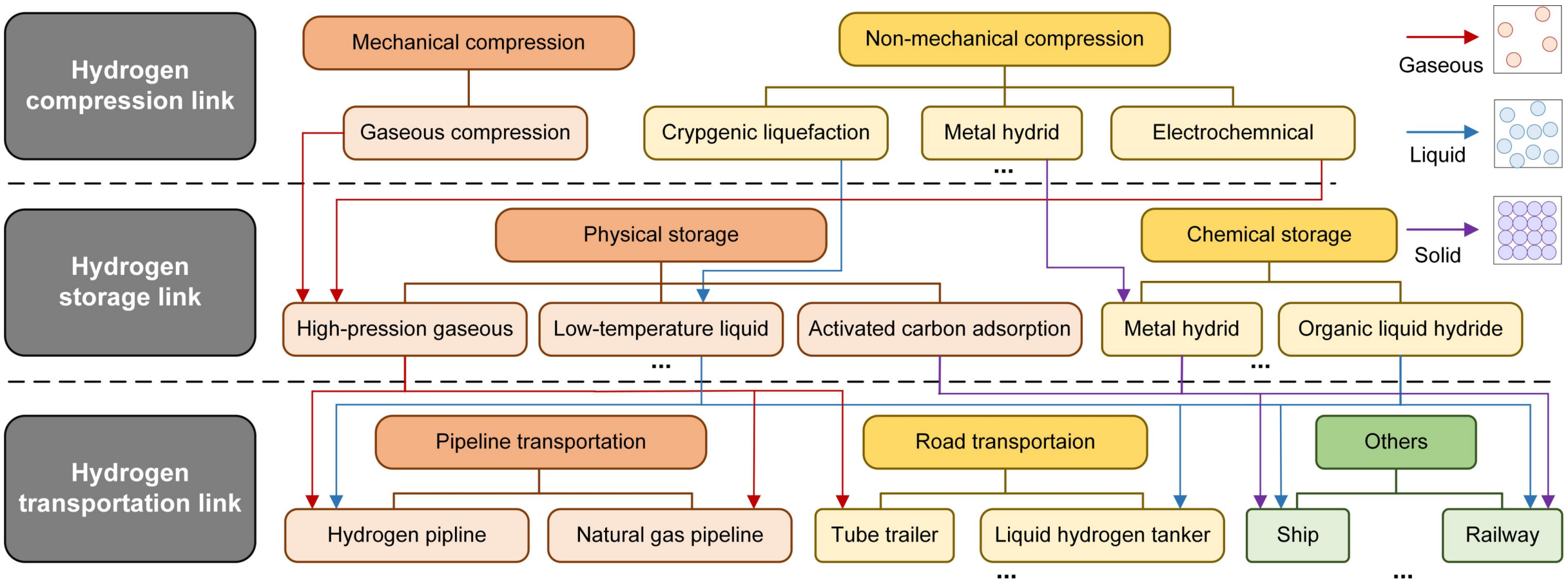

**Fig. 4.** Classification of technologies in hydrogen compression, storage and transportation link

## *2.4 Hydrogen transportation link*

Hydrogen transportation serves as a bridge in the HEC, connecting isolated hydrogen production, storage, or application sites to achieve optimal allocation of hydrogen energy across space. The mode of transportation is usually closely related to the hydrogen form, distance, and scale. Common modes include gaseous hydrogen tube trailers, liquid hydrogen trucks, pipelines, ships, and railways [48-49]. Liquid hydrogen trucks offer a cost-effective and high-capacity solution for long-distance transportation.

In addition, liquid hydrogen can be transported over long distances by ships and railways, although safety concerns regarding leakage must be considered [50]. Compared with liquid hydrogen trucks, solid hydrogen storage vehicles are considered safer and can be transported at normal temperatures and pressures. However, because of technical and cost constraints, only small-scale demonstration applications have been realized [51]. The main mode of large-scale hydrogen transportation over long distances is pipeline transportation of gaseous hydrogen, but the infrastructure is costly and has long construction periods. In the world, approximately 5,000 km of existing hydrogen pipelines and 1 million km of natural gas pipelines are in operation [35]. Therefore, blending hydrogen with natural gas to utilize existing pipeline infrastructure is considered a promising transition method to enable a low-cost, scaled, and continuous hydrogen energy supply. The gaseous hydrogen tube trailer with a small single-vehicle capacity is a mature technology that is more economical for short distances [49]. In addition, unlike closed pipeline networks, gaseous tube trailers offer the flexibility to transport mobile hydrogen storage tanks (HSTs) from production and application sites [52]. Hydrogen transportation can address the spatial imbalance between the hydrogen load and demand, contributing to the optimal allocation of energy resources across space.

## *2.5 Hydrogen application link*

Hydrogen application is the final stage of HEC. With various conversion and application scenarios, hydrogen will contribute to the deep decarbonization of chemical, industrial, energy, traffic, and other sectors in the future, as shown in Fig. 5. In 2022, global hydrogen consumption reached 95 million tons, with 41 million tons used in petroleum refining and 53 million tons in synthetic ammonia, synthetic methane, and steel smelting [35]. At this stage, hydrogen is used mainly for chemical feedstock production, which accounts for approximately 99% of the total hydrogen consumption. While hydrogen consumption in traffic is currently negligible, it is projected to exceed 30% of the total energy consumption of traffic by 2050 [53]. Hydrogen can provide power for various modes of traffic sector, including road, rail, marine, and air transportation. The prevailing application scenarios are HFCVs with hydrogen refueling stations (HRSs) [54]. FC technology can also be applied to energy systems for distributed generation or cogeneration, while HTs and hydrogen-fired boilers (HBs) can also be utilized for energy supply [55]. Moreover, hydrogen can react with $CO_2$ captured by CCS to produce clean natural gas [56]. As a flexible and efficient energy carrier, hydrogen can realize the interconnection and complementarity of various energy forms, such as electric power, heat, and natural gas, through diverse application scenarios.

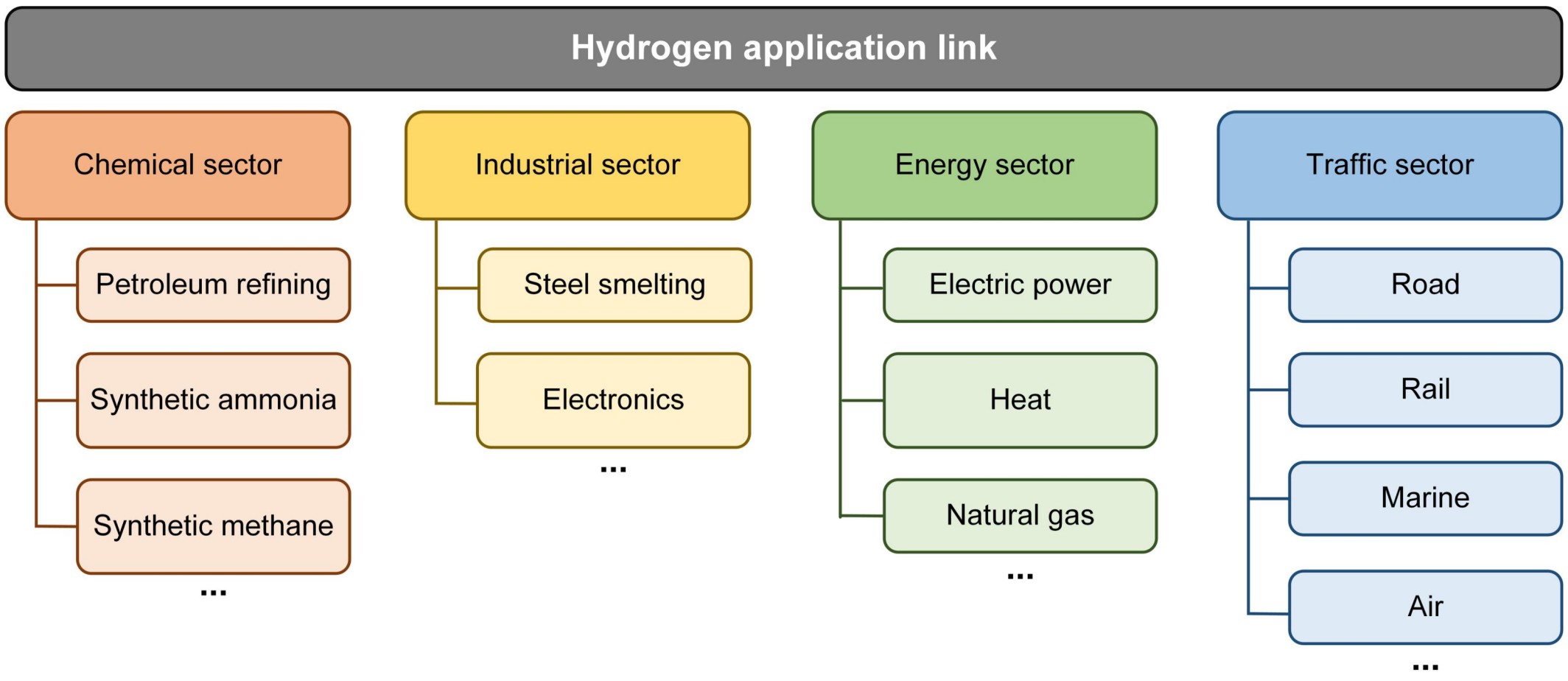

**Fig. 5.** Classification of technologies in hydrogen application link

## *2.6 Relationships between the links of the HEC*

The diverse links in the complete HEC are closely interrelated and interact with each other, and the fundamental composition is shown in Fig. 6. Hydrogen production is the starting point of HEC and is related to the operation of all subsequent links. It converts feedstocks such as fossil fuel, electricity, and water into hydrogen. Compression is necessary to meet the requirements of storage and transportation links and directly affects storage efficiency. The hydrogen storage link enables hydrogen transfer over time, and its technology choice determines the mode of hydrogen transportation. For example, pipelines or tube trailers are generally chosen for gaseous storage, while liquid hydrogen trucks or ships are used for liquid storage. The production and application links determine the charging or discharging of the HS. Transportation link connects remote terminals to achieve link synergy and hydrogen transfer across space. The application link is the end point of the HEC. The other links synergize to meet the hydrogen demand of the application end. Moreover, the application end influences the operation and configuration of the other links.

It is worth noting that the conversion relationship of the energy flows strengthens the coupling between the links of HEC in the MES, contributing to synergistic planning of HEC-MES and optimizing overall investment.

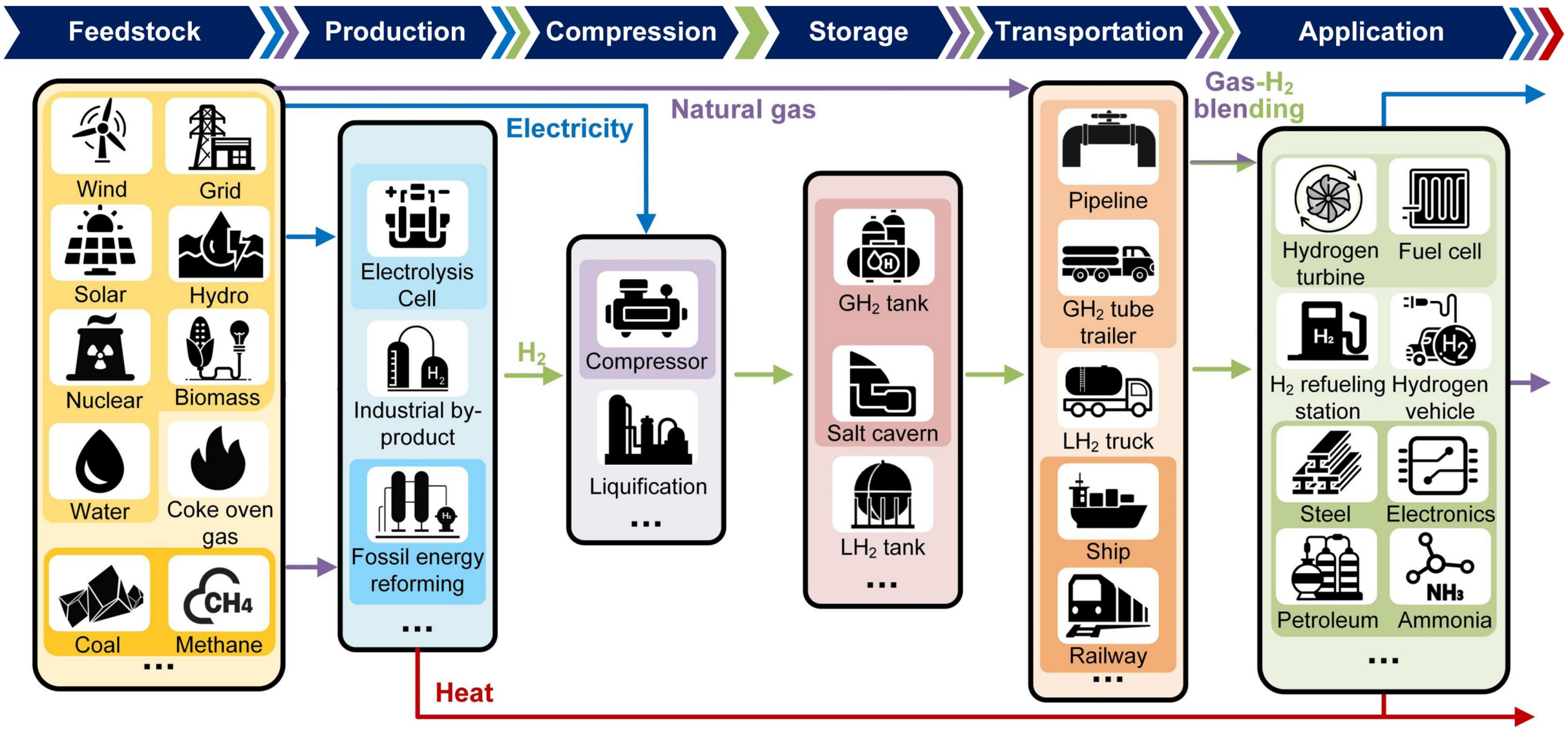

**Fig. 6.** Composition of the HEC (the technology is simply classified by the shade of the background color)

# 3. HEC planning in HEC-MES

This section focuses on current research on HES-MES planning. We review the link composition and technology deployment of HECs in MES planning research in recent years and summarize them in Table 1. Fig. 7 presents the steps of the paper review. First, we used search engines and keywords to search relevant literature published in recent years. Then, we filtered the data based on the impact factor, source, and type. Finally, we selected 72 papers for detailed analysis based on title, abstract, and content.

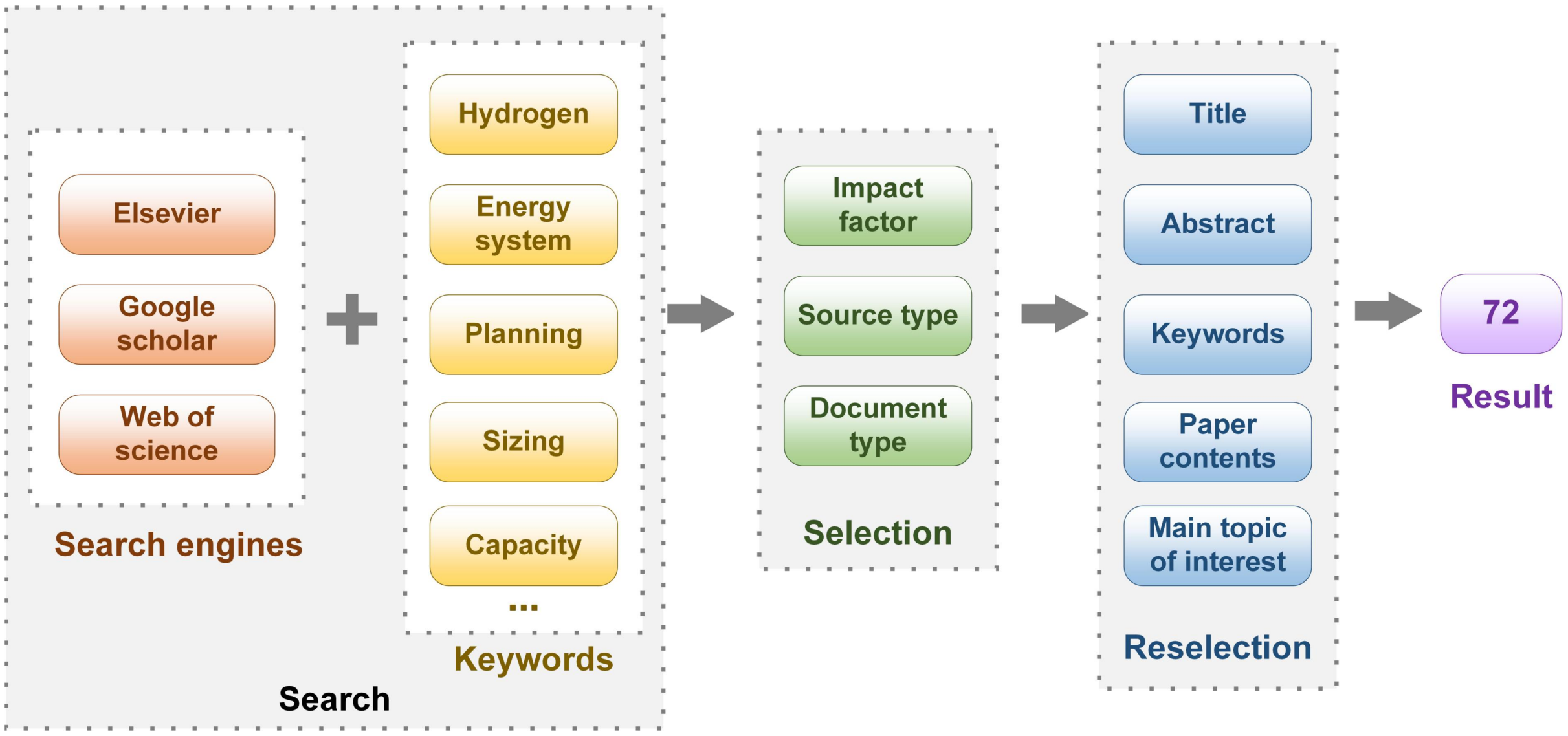

**Fig. 7.** Screening process of relevant literature

**Table 1 -** Composition of HEC and technology deployment in the literature

| Ref. | Production | Compression | Storage | Transportation | Application | Form of HEC |
|---|---|---|---|---|---|---|
| [22] | EC | - | SC | - | FC+D | EC-SC-FC/D |
| [23] | EC | COP (·) | HST | - | FC | EC-COP-HST-FC |
| [24] | EC | - | HST | - | FC | EC-HST-FC |
| [57] | EC | - | HST | - | FC+Sell | EC-HST-FC/Sell |
| [58] | EC | - | * | - | FC+D+Sell | EC-S-FC/D/Sell |
| [59] | EC | * (·, E) | HST (SEA) | - | FC+Sell | EC-C-HST-FC/Sell |
| [60] | EC+P | - | HST (SEA) | *(·, CT) | FC+HV+sell | EC-HST-T-FC/HV; EC/P-HST(SEA)-T-Sell |
| [61] | EC | - | * (SEA) | - | GT+D+sell | EC-S-GT/D/Sell |
| [62] | EC | * (·, E) | HST(SEA) | - | FC+sell | EC-C-HST-FC/Sell |
| [63] | EC | - | - | - | HT | EC-HT |
| [64] | EC | COP (·, LS) | HST (SEA) | PT (·, LS) | FC | EC-COP-HST-PT-FC |
| [65] | EC | * (·, E) | SC (SEA) | PT (·)+RT (·) | HRS+HFCV+H2P(·) | EC-C-HRS-HFCV/H2P EC-C-SC-PT/RT |
| [66] | EC | COP | HST | RT (trailer) | HRS+HFCV | EC-COP-HS-HRS-HFCV/RT |
| [67] | EC | COP+EXP | *(SEA) | PT (·) + RT (trailer) | D+Sell | EC-COP-S-EXP-D; EC-COP-S-EXP-COP-RT-Sell |
| [68] | EC+SMR(CCS)+IM | COP | HST | PT | FC+HT+D+curtailment+EX | IM/EC/SMR-COP-HST-FC/HT/D/Curtailment/EX |
| [69] | EC | COP | HST | - | HRS+HFCV | EC-COP-HST-HRS-HFCV |
| [70] | EC | - | HST (SEA) | PT | H2P+D | EC-HST-PT-H2P/D |
| [71] | EC+F2H | - | *(SEA+MOB) | RT | D | EC-S-RT-HS-D; F2H-D |
| [72] | EC | - | HST | - | HRS+HFCV | EC-HST-HRS-HFCV |
| [73] | EC | - | HST | - | FC | EC-HST-FC |
| [74] | EC | COP+DP | HST (SEA) | - | D | EC-COP-HST-DP-D |
| [75] | EC | COP (·) | SC | PT | FC+D | EC-COP-HS-FC/D |
| [76] | EC | - | - | - | Sell | EC-Sell |
| [77] | EC | COP | HST+SC (SEA) | PT | D | EC-COP-HST/SC-PT-D |
| [78] | EC+SMR | COP+L | HST (SEA+MOB) | PT+RT (truck) | D | EC/SMR-COP/L-HST-PT/RT-D |
| [79] | EC | - | - | PT | HRS+HV | EC-PT-HRS-HV |
| [80] | EC | COP+L (·) | * | RT (trailer) | HRS+D | EC-COP/L-HS-RT-HRS-D EC-COP/L-D |
| [81] | EC | COP (·, LS) | HST (SEA) | - | FC+D | EC-COP-HST-FC/D |
| [82] | EC+P | COP (·, E) | HST | PT (·)+RT (·) | FC | EC/P-COP-HST-PT/RT-FC |
| [83] | EC | COP+PR (·) | HST (SEA) | PT | D | EC-COP-HST-PT-PR-D |
| [84] | EC | - | HST | PT | HRS+HFCV | EC-HST-PT-HRS-HFCV |
| [85] | EC | PR (·,LS) + DP (·,E) | HST (SEA) | - | HRS+HFCV | EC-HST-DP-HRS-HFCV |
| [86] | EC | - | HST | - | FC | EC-HST-FC |
| [87] | EC | COP | HST | - | FC | EC-COP-HST-FC |
| [88] | EC | - | HST(SEA) | - | FC+SM | EC-HST-FC/SM |
| [89] | EC+IBH | COP (·) | HST | - | HRS+FC+HV+HFCV | EC-COP-HST-HRS-FC/HV/HFCV; IBH-HST-HRS-FC/HV/HFCV |
| [90] | EC | - | HST(SEA) | PT (·) | FC+D | EC-PT-HST-FC/D |
| [91] | EC | - | HST | - | FC | EC-HST-FC |
| [92] | EC | COP (·) | HST(MOB) | - | FC+D | EC-COP-HST-COP-HST (MOB)-FC/D |
| [93] | EC | - | HST(SEA) | PT | HCHP+HB | EC-HST-PT-HCHP/HB |
| [94] | EC | COP (·) | HST | PT (·) | FC+HRS+HFCV+D | EC-COP-HST-PT-FC/D; EC-COP-HST-PT-HRS/HFCV |
| [95] | EC | - | HST(SEA) | - | FC+D | EC-HST-FC/D |
| [96] | EC+P | COP | * | - | FC | EC/P-COP-S-FC |
| [97] | EC | COP (·) | HST | - | HFCV | EC-COP-HST-HFCV |
| [98] | EC | COP+DP+PC | HST (LPT/HPT) | - | FC+D | EC-LPT-FC EC-LPT-COP-HPT-DP/PC-D |
| [99] | EC | COP | HST | PT (·) | Sell | EC-COP-HST-Sell |
| [100] | EC | COP | HST | - | FC+HV | EC-COP-HST-FC/HV |
| [101] | EC | - | HST (SEA) | - | FC | EC-HST-FC |
| [102] | EC | - | HST | - | FC | EC-HST-FC |
| [103] | EC | - | HST | - | FC | EC-HST-FC |
| [104] | EC | COP | HST | - | FC | EC-COP-HST-FC |
| [105] | EC | - | HST | - | FC | EC-HST-FC |
| [106] | EC | - | HST+SC (SEA) | - | FC | EC-HST/SC-FC |
| [107] | EC+P | - | HST | - | FC+HV | EC/P-HST-FC/HV |
| [108] | EC | - | HST | - | FC | EC-HST-FC |
| [109] | EC+IM | COP (·, E) | HST | PT (·)+RT (·) | FC+GT+D | EC-COP-HST-FC/GT/D; IM-PT/RT-FC/GT/D |
| [110] | EC+SMR(CCS) | COP | HST+SC (SEA) | PT | FC+D | EC-SMR -HST/SC |
| [111] | EC | - | - | - | D+Sell | EC-D/Sell |
| [112] | EC | COP | HST | RT (trailer) | HRS (·)+SA (·) | EC-COP-HST-RT-HRS EC-COP-HST-SA |
| [113] | EC | - | HST+SC (SEA) | - | FC | EC-HST/SC-FC |
| [114] | EC | - | - | - | D+Sell | EC-D/Sell |
| [115] | EC | - | HST (SEA) | - | FC | EC-HST-FC |
| [116] | EC | COP (·) | HST | PT (·)+RT (·) | HT+Sell | EC-COP-HST-PT/RT-HT/Sell |
| [117] | EC | - | HST | - | FC | EC-HST-FC |
| [118] | EC | - | HST | - | FC+HFCV | EC-HST-FC/HFCV |
| [119] | EC | - | HST (SEA) | - | D | EC-HST-D |
| [120] | EC | - | HST | - | FC | EC-HST-FC |
| [121] | EC | - | HST | - | FC | EC-HST-FC |
| [122] | EC | - | HST | - | FC | EC-HST-FC |
| [123] | EC | - | HST | - | FC | EC-HST-FC |
| [124] | EC | COP (·, E) | HST | - | FC+SA | EC-COP-HST-FC/SA |
| [125] | EC | - | HST | - | FC | EC-HST-FC |

*, considered, but technology not specified; ·, mentioned only, no planning considered; T, transportation; C, compression; S, storage; LS, considered loss; CT, considered cost; E, considered electricity consumption; COP, compressor; SC, salt cavern; PT, pipeline transportation; RT, road transportation; D, demand; GT, gas turbine; SMR, steam methane reforming; IM, import; EX, export; SEA, seasonal; L, liquefaction; P, purchase, MOB, mobile; H2P, hydrogen-to-power; HV, hydrogen vehicle; DP, dispenser; EXP, expander; F2H, fossil to hydrogen; PR, pressure regulation; SM, synthetic methane; IBH, industrial by-products for hydrogen production; HCHP, hydrogen combined heat and power; LPT, low pessure tank; HPT, high pressure tank; PC, pre-cooler; SA, synthetic ammonia.

### 3.1 Planning for hydrogen production technology

Table 1 exhibits that production is an essential component of HEC-MES, which is the foundation for HEC. All of the reviewed papers applied EC technology for hydrogen production. Other methods consider hydrogen production from fossil energy sources [68,71,78,110] and industrial by-products [89]. Considering the limitations of renewable energy output, hydrogen production from ECs is volatile. Therefore, some of the literature considers the purchase of hydrogen from others to compensate for the shortfall of hydrogen produced by ECs [60,82,96,107], while others import hydrogen in the form of ammonia [109]. As an essential technology in HEC-MES planning for a low-carbon future, the EC contributes to renewable energy consumption, increases safety and stability, and decarbonizes the energy, industry, and traffic sectors [58].

The role of ECs in MESs has been analyzed in detail by many researchers, who have concluded that the allocation of EC capacity in energy system planning is closely related to the distribution of renewable energy resources. Reference [68] modeled planned investments in 2050 for an electric-hydrogen energy system in Texas. The results indicate that ECs allow energy systems to integrate more renewable energy with fewer energy storage configurations. Literature [88] provided annual high-resolution planning for South American energy systems' transition to 100% renewable energy systems. In South America, the electricity consumption of green hydrogen production mainly relies on PVs. The ECs preferred to invest in solar-rich countries such as Chile. As hydrogen production increased, the average cost of electricity in South America's energy systems decreased. That is because ECs replace other flexibility sources (e.g., gas turbines (GTs)) and reduce the abandonment of renewable energy. Xuejie Wang et al. [96] proposed a combined wind-storage system that incorporates hydrogen energy using distributed robust optimization (DRO) and revenue allocation based on wind energy uncertainty. The example simulations showed that considering wind energy fluctuations helps to improve the utilization efficiency of wind energy, effectively reducing the hydrogen production costs and increasing the system's overall profitability. Literature [119] proposed a synergistic planning method for hydrogen production systems by combining different types of ECs to achieve economical and efficient operation.

Although EC technology has various strengths as previously mentioned, the investment cost limits its large-scale application. The cost of electrolysis for hydrogen production exceeds that of fossil energy catalytic reforming by more than five times [126]. Research [57] analyzed multiple actors' interests in electric-hydrogen energy systems planning. The article concluded that it is uneconomical to purchase electricity for hydrogen production regardless of the efficiency of ECs. Utilizing curtailed renewable energy for hydrogen production can lower purchased electricity costs. However, heat loss in the electricity-to-hydrogen process results in low energy conversion efficiency. Therefore, some of the literature considers waste heat utilization of ECs [24,61,64,74,105]. In addition to reducing power purchase costs and improving energy conversion efficiency, some researchers have introduced carbon emission costs to enhance the competitive advantage of ECs. Reference [68] modeled the effect of $CO_2$ price and hydrogen demand on the hydrogen production methods' selection. When the $CO_2$ price was 0 $/t, the energy system chose to achieve economic optimality by utilizing only steam methane reforming (SMR) for hydrogen production. With the increase in $CO_2$ prices, EC technology was gradually replacing SMR technology as the primary method. When the $CO_2$ price exceeded $90/t, SMR with CCS technology replaced the remaining SMR technology because renewable energy installation limited the increased capacity of ECs. At the same $CO_2$ price, the proportion of hydrogen produced from SMR with CCS technology gradually increased as the hydrogen demand increased. That is because SMR with CCS technology becomes more advantageous as the demand for hydrogen increases while considering the cost-efficiency of hydrogen production.

During water electrolysis, the by-product oxygen comes out alongside hydrogen [127]. The revenue from oxygen sales can offset some of the costs, and considering it during EC planning can provide a more comprehensive economic simulation [128]. Literature [99] finely described the dynamics of the flow and pressure of oxygen during the production, compression, and storage processes and accurately simulated revenues from oxygen sales. However, only a small amount of literature has considered the benefits derived from oxygen as a by-product of hydrogen production [129]. To accurately model the impact of ECs on HEC-MES planning, it is crucial to consider not only hydrogen but also by-product benefits such as waste heat utilization and oxygen sales, which remains an area that requires further study.

### *3.2 Planning for hydrogen compression technology*

As shown in Table 1, more than half of the investigated literature on HEC-MES planning neglected to plan for the hydrogen compression link. In this type of literature, compression is usually incorporated into the production link or storage link in the form of efficiencies. In the literature considering the hydrogen compression link, the compressor is the most commonly used compression technology for pressurizing low-pressure gaseous hydrogen into high-pressure gaseous hydrogen. However, some of the papers only mention compressors as a technology without conducting a planning analysis [23,75,89,92,94,97,116], while others only consider compressor-induced losses [64,81] and power consumption [82,109,124]. Similarly, the reference [59,62,65], which does not specify the compression technique, considers only the power consumption caused in the compression link. Many researchers ignore the location, capacity sizing, and investment costs of the compressors when planning to simplify the complexity of the model.

A simplified model of the compressor is used in general energy planning, considering only a fixed power consumption factor [87] or a fixed overall efficiency [130] for a single type. In addition to electricity, gas can also drive compressors, so some researchers have considered the gas consumption of the compression link [77]. According to the U.S. Department of Energy, the energy consumption of actual hydrogen compression is affected by a combination of the compressor flow rate, pressure ratio, efficiency, and other factors [131]. Therefore, a significant deviation remains between the simulation results of the above literature with a simplified compressor model and the actual situation. Literature [69] developed a refined compressor power consumption model considering the active power consumption rate, operating pressure level, and hydrogen dissipation rate. Reference [96] proposed a compressor power consumption model that integrates temperature, efficiency, pressure ratio, and flow rate. In practice, various application scenarios, such as hydrogen storage, seasonal hydrogen storage, and tube trailers, require different hydrogen pressure levels and different types of compressors to match. Jianxin Lin et al. [67] proposed a compressor model that can adapt to different hydrogen pressure demands by combining the effects of temperature, flow rate, and efficiency. The model allowed the overall efficiency of the compressor to be adjusted with different levels of input and output pressure. The results show that considering various compressor types can decrease the total invested compressor capacity.

In addition to the most commonly used gas compressors, energy system planning incorporates other compression techniques. Literature [78,80] considered the application of liquefaction techniques. However, the above studies lack a detailed parametric or modeling distinction between hydrogen compression and liquefaction. Furthermore, compressed hydrogen needs to be readjusted to the appropriate pressure level using devices such as expanders [67], distributors [74,85], or regulators [83] during the use of the HRS or pipelines. Research [98] also considered the planning of pre-coolers to cope with the temperature rise of the hydrogen refueling process. The HEC-MES planning study lacks refined modeling and detailed analysis of the compression link.

### *3.3 Planning for hydrogen storage technology*

Seasonal HS is a crucial approach for achieving large-scale, long-duration energy transitions. Therefore, the hydrogen storage link is currently the focus of HEC research for HEC-MES planning. Only [63,76,79,111,114] of the relevant literature surveys neglected the hydrogen storage link, and some of the others did not specify the HS technology used [58,61,67,71,80,96]. Moreover, research shows that HSTs are the most commonly used hydrogen storage technology. They are usually coupled with ECs to realize on-site storage and further enhance the energy system flexibility. Literature [98] considered low- and high-pressure HSTs to meet the hydrogen pressure requirements for FC and hydrogen loads, respectively. Reference [66] considered the uncertainty of wind energy and hydrogen demand for hydrogen infrastructure planning. The simulation results show that the investment cost of the storage battery is much greater than the obtainable benefits. Therefore, only investing in HSTs for a wind fluctuation regulation policy can reduce the total investment cost. Soheil Mohseni et al. [94] conducted a validation analysis of the financial viability of HSTs. Although it is less expensive to utilize lead-acid and lithium batteries instead of HSTs for planning, both types of energy storage are environmentally harmful. Considering the cost of environmental treatment, the total cost will be greater than that of the HST planning scenario. In addition, the application of HSTs avoids the idling of ECs when the hydrogen load is zero, allowing for maximum utilization of the ECs' potential.

Researchers utilize HSTs in energy systems for short-term and seasonal HS research. The difference in parameters between them is mainly in the cost. For instance, in the literature [61], the short-term and seasonal HST costs are \$15/kWh and \$0.91/kWh, with a difference of more than ten times. In addition, short-term HSTs are regulated by charging and discharging hydrogen within a day, whereas seasonal HSTs are regulated between days with lower charge cycles per year. Guangsheng Pan et al. [81] conducted a collaborative planning analysis of HSTs across different time scales. This paper sets the state of charge of the short-term HST to be the same at 0:00 and 24:00, with only one operation state (charge/discharge) in the seasonal HST each day. The challenge in modeling seasonal HSTs is to simplify the complexity of planning on long-time scales while simulating the variation in hydrogen stored between seasons. Conventional planning methods based on typical days with time discontinuities cannot simulate seasonal hydrogen storage. To address this issue, literature [62] proposed a new mixed integer linear programming (MILP) planning method based on the characteristics of different devices in the energy system. This approach can reduce problem complexity while realizing hourly high-resolution long-term planning.

In HEC-MES, hydrogen can also be stored in underground geologic formations such as depleted oil fields, aquifers, and salt caverns (SCs). SCs are considered the most suitable underground storage media for HS because of their gas tightness and lack of reaction with hydrogen [132]. Compared to HSTs, SCs also have the advantages of a small footprint aboveground, low price, and high safety [133]. Study [22] analyzed the impact of SC costs on an energy system using the example of a future European zero-carbon emission power system. The article argued that the lower the SC costs, the better the energy system meets the demand for flexibility and large-scale energy storage. In addition, the availability of SCs becomes an important internal driver for the storage scale, taking into account natural constraints. For example, reference [113] chose Anhui Province as a research case because of its abundant SC resources. Hydrogen stored in SCs is typically used for seasonal storage in energy systems to regulate the imbalance between energy supply and demand over time. Vincent Oldenbroek et al. [75] examined the seasonal variation in hydrogen storage in an SC. Owing to the summer concentration of PV, the seasonal renewable energy output and the load curve are more mismatched as the PV proportion increases. A lower match results in a more significant regulation effect of SCs on the energy system, requiring more new installed capacity.

In HEC-MES planning, several articles consider multiple energy storage methods. For instance, literature [97] conducted synergistic planning of hydrogen, electricity, heat, and cold storage devices. The optimal allocation for multiple energy storage is crucial for coordinating heterogeneous energy flexibility and smoothing renewable energy volatility. It has many benefits, including compensating for power imbalances on multi-time scales [90] and reducing system investment costs [82]. In addition, HS also plays a role in HEC-MES planning for cross-space energy transfer due to its mobility, which is detailed in the next section.

### 3.4 Planning for hydrogen transportation technology

The construction of a hydrogen transportation network can liberate the links of the HEC from regional limitations. However, the hydrogen transportation link is widely ignored in current research on HEC-MES planning since most of these studies focus on a single region or energy system. To simplify the complexity of the model, researchers usually assume unrestricted intra-regional hydrogen transfer. In other words, it is assumed that the transfer of hydrogen between hydrogen production and hydrogen consumption ends by any means without considering losses and costs [82]. Additionally, some multi-region energy systems overlook inter-regional hydrogen transfer planning, assuming hydrogen balances within the region. Some literature mentioned only the transportation link without addressing the planning of specific equipment and considered only the costs [60] or losses [64] during transportation. In HEC-MES planning, pipeline and road transportation are two commonly used types of hydrogen transportation.

Some hydrogen pipeline transportation models typically consider capacity limitations and inter-regional exchange balance constraints [68]. These limit the maximum flow and ensure that the hydrogen energy transferred from one region to the other equals the sum of the flows of the pipelines connected to it. Some employ the Weymouth model, where the gas steady-state flow rate is determined based on the pressure at both ends of the pipeline [134]. In addition, the model usually includes the hydrogen balance and upper/lower bound constraints on the flow rate and pressure [84]. Although this model type is accurate, it is generally only used for refined modeling of small energy systems at the distribution network scale because of its non-convexity, computational complexity, and solution difficulty. In addition, retrofitting natural gas pipelines for blended hydrogen transportation is a more economically competitive option than adding new hydrogen pipelines [77]. Literature [110] proposed a pipeline transport model for gas-hydrogen blending based on steady-state gas flow and noted that ignoring blended hydrogen transportation is prone to leading the planning results into sub-optimality.

In a broad sense, hydrogen transportation can be a mobile HS. In a narrow sense, mobile HSs usually refer to the utilization of vehicles such as tube trailers or trucks to transport HSTs from production to consumption ends. Reference [92] utilized hydrogen-fueled vehicles as carriers for HSTs and generated power through FCs consuming hydrogen from mobile HSTs to supplement the power for building clusters. Mobile HS is a specific form of hydrogen road transportation. Based on the mobility and seasonal balance characteristics of HSs, literature [71] proposed a mobile seasonal HS to simulate the inter-regional road transportation of hydrogen.

### *3.5 Planning for hydrogen application technology*

As the most downstream part of the HEC, the hydrogen application link is as essential as the hydrogen production link in HEC-MES. This article describes the diverse applications of hydrogen in Section 2.5. Table 1 shows that the forms of hydrogen applications in HEC-MES can be categorized as outside or inside the energy system.

Hydrogen demand usually represents hydrogen applications outside the energy system, with data based on estimates and forecasts. Literature [58] estimated the hydrogen demand in 10 cities based on the production of chemical products (e.g., iron and steel, oil refining, and ammonia) and primary energy consumption (e.g., diesel, gasoline, and natural gas). Espen Flo Bødal et al. [68] used the NERL projected traffic hydrogen demand for 2050 as a baseline to perform a planning analysis and excluded industrial hydrogen demand. This is because on-site hydrogen production supplies much of the industrial hydrogen demand. Reference [95] forecasted the hydrogen demand in 2050 based on data from 2030 with an average annual growth rate of 1%. The data includes the hydrogen demand for traffic, building, industry, and other sectors. Literature [112] has assumed that all produced hydrogen is used to meet the hydrogen demand for ammonia synthesis and HRSs. Moreover, the proceeds of hydrogen sales to satisfy external hydrogen demand can offset the investment cost of hydrogen equipment and enhance the economics of HEC-MES [76]. However, the above research ignored the temporal characteristics of hydrogen demand and distributed the annual hydrogen demand evenly each hour. This approach is inconsistent with realistic scenarios and is unable to simulate more accurate planning and operation of hydrogen energy equipment. Hydrogen demand uncertainty can be captured by building uncertainty sets [61,78], introducing stochastic volatility rates [74], and utilizing scenario analysis techniques [66], such as combining Monte Carlo (MC) methods and density-based clustering methods. However, these methods do not accurately characterize fluctuations in hydrogen demand, so accurate modeling of hydrogen demand uncertainty still needs to be addressed.

There is a growing global consensus that hydrogen will replace conventional fuels as the primary means of decarbonizing the traffic sector in the future [135]. Hydrogen-fueled vehicles are expected to increase significantly, especially in the United States, Germany, Japan, and China [136]. Researchers are increasingly introducing hydrogen-fueled vehicles in HEC-MES planning to visualize the hydrogen demand in the traffic sector. These usually involve the configurations of HRSs and traffic networks. Literature [79] proposed a multi-network collaborative planning model to cope with the variability of renewable energy and hydrogen-fueled vehicle loads by optimizing investment decisions for hydrogen pipelines, HRSs, ECs, and PVs. Yuchen Dong et al. [89] introduced HFCVs in hydrogen-based microgrid planning and verified that their emergency response can effectively improve grid resilience. In addition, combining hydrogen with energy systems contributes to decarbonization in the chemical sector. Hydrogen energy combined with CCS technology can help to achieve more cost-effective and low-carbon methane production [88]. Research [124] proposed a stand-alone renewable energy system planning model considering ammonia, where optimal economic benefits were obtained through rational planning of renewable energy, ECs, FCs, HSTs, and synthetic ammonia equipment. The impact of technologies such as ammonia energy storage and ammonia-fired generation on MESs is a valuable future research direction.

The primary use of hydrogen within energy systems is to convert it into electrical or heat energy through devices such as FCs [62], HTs [63], and HBs [93]. Due to the immaturity of HT technology for pure hydrogen, some researchers blend hydrogen into natural gas and utilize a hydrogen-blended GT for cogeneration [61]. Literature [103] utilized a natural gas-hydrogen mixture with a 15% hydrogen volume concentration for cogeneration units and boilers. The volume of the gas mixture needs to change according to the hydrogen percentage. Using hydrogen for clean electricity and heat supplies can reduce carbon in energy systems. However, because of the low energy conversion efficiency of the electricity-hydrogen-electricity process, hydrogen-to-power (H2P) technology contributes little to replenishing the energy deficit and improving flexibility [13]. Therefore, H2P technologies with low efficiency for balancing supply and demand in energy systems are unpopular. According to the literature [75], the installed FC capacity gradually decreased as the hydrogen demand increased. Coincidentally, research [95] presented that when the hydrogen demand is high, its growth causes a reduction in the total installed capacity of FCs. Furthermore, the selection of FCs favored more efficient solid oxide fuel cells (SOFCs) instead of more flexible and less expensive proton exchange membrane fuel cells (PEMFCs). Efficiency gains and cost reductions in future energy conversion equipment can change this outcome.

## *3.6 Forms of HEC in HEC-MES*

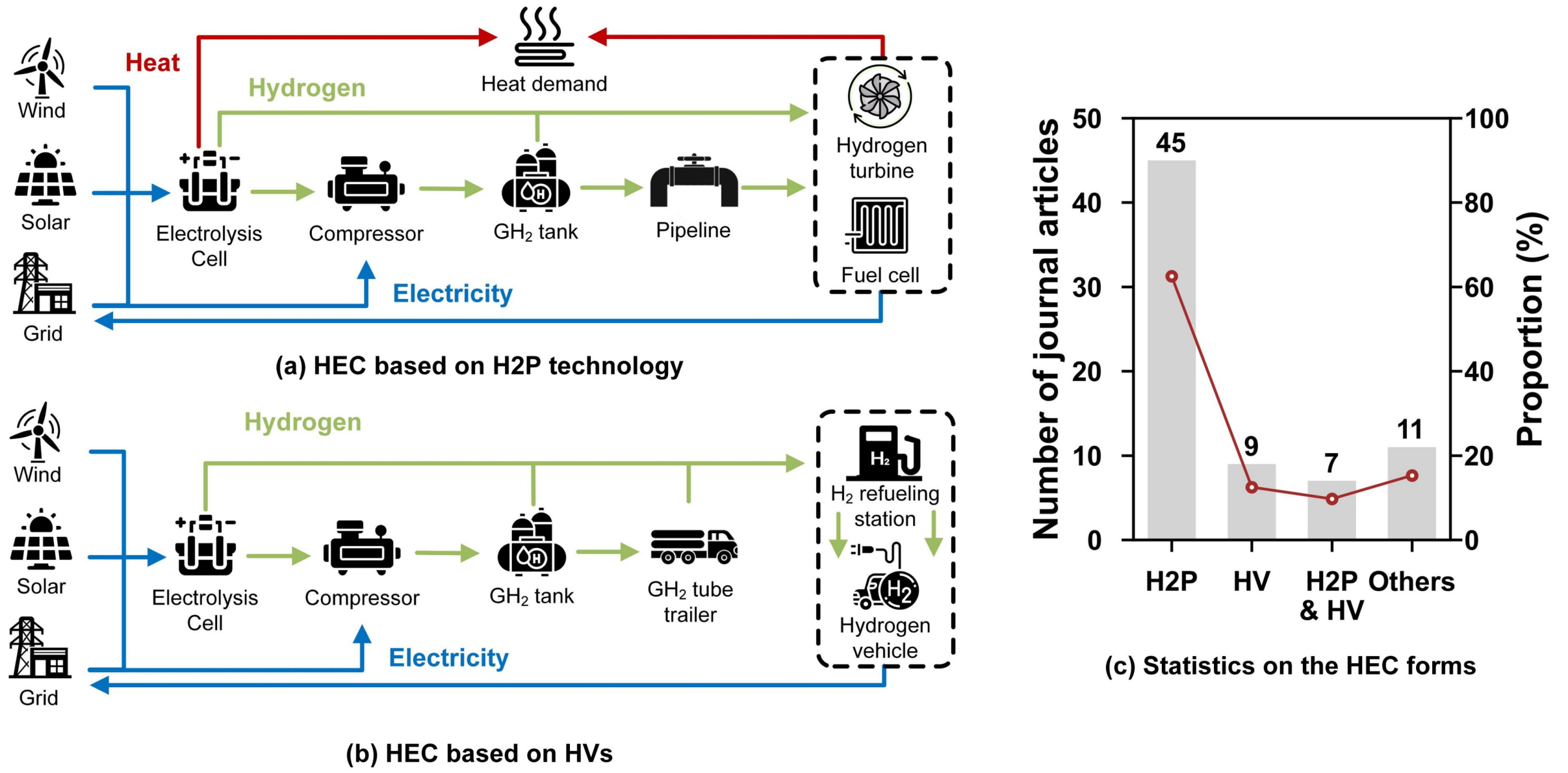

**Fig. 8.** Typical forms of HEC and statistical results

Table 1 presents statistics on the various forms of HEC in HEC-MES resulting from different combinations of links and devices. This paper summarizes two typical forms of HECs based on hydrogen application scenarios: 1) HEC based on H2P technology, as shown in Fig. 8(a). 2) HEC based on hydrogen vehicles (HVs), as shown in Fig. 8(b). Figure. 8(c) illustrates the number and percentage of literature on typical types of HECs. The most common of the first forms consists of an EC, compressor, HST, and FC combined as a single unit to smooth out the volatility of renewable energy [86]. This HEC for power and heat supply contributes to carbon reduction and an increase in energy efficiency [91]. The second form usually involves a combination of HRSs and HVs [79]. This coupling of energy and traffic networks through HVs fully utilizes multi-network synergy and promotes the decarbonization of the traffic sector [66].

Figure. 9 illustrates one form of a complete HEC in a MES. The process is as follows: With surplus renewable energy generation, ECs produce green hydrogen to reduce wind and solar curtailment. The hydrogen is then compressed using a compressor to a reasonable pressure level and stored in HSTs. With a lack of energy supply, the stored hydrogen can be used directly in FCs for clean cogeneration and in synergy with CCS for methane synthesis. On the other hand, hydrogen can also be transferred through pipelines or tube trailers to remote hydrogen-consuming terminals for application. The complete hydrogen energy chain facilitates energy transfer across time and space, converting surplus renewable energy generation into flexible and controllable clean energy. This enhances the bi-directional regulation capability of MESs, improves efficiency and flexibility, and reduces carbon emissions.

However, few studies in the researched literature have analyzed complete HEC planning that includes all the links. HEC links are closely connected and interact with each other. Incomplete HEC will hinder the integration of hydrogen energy and MESs, limiting advantages such as economy, flexibility, and carbon reduction. The results in the study [24] indicate that neglecting the hydrogen storage link leads to a rise in costs, carbon emissions, and the installed capacity of other energy devices, resulting in a decrease in energy efficiency. Complete HEC planning research can help to optimize hydrogen energy allocation on temporal and spatial scales and promote diverse hydrogen applications. In addition, the challenges of model complexity and solution difficulty for refined complete HEC modeling require further research.

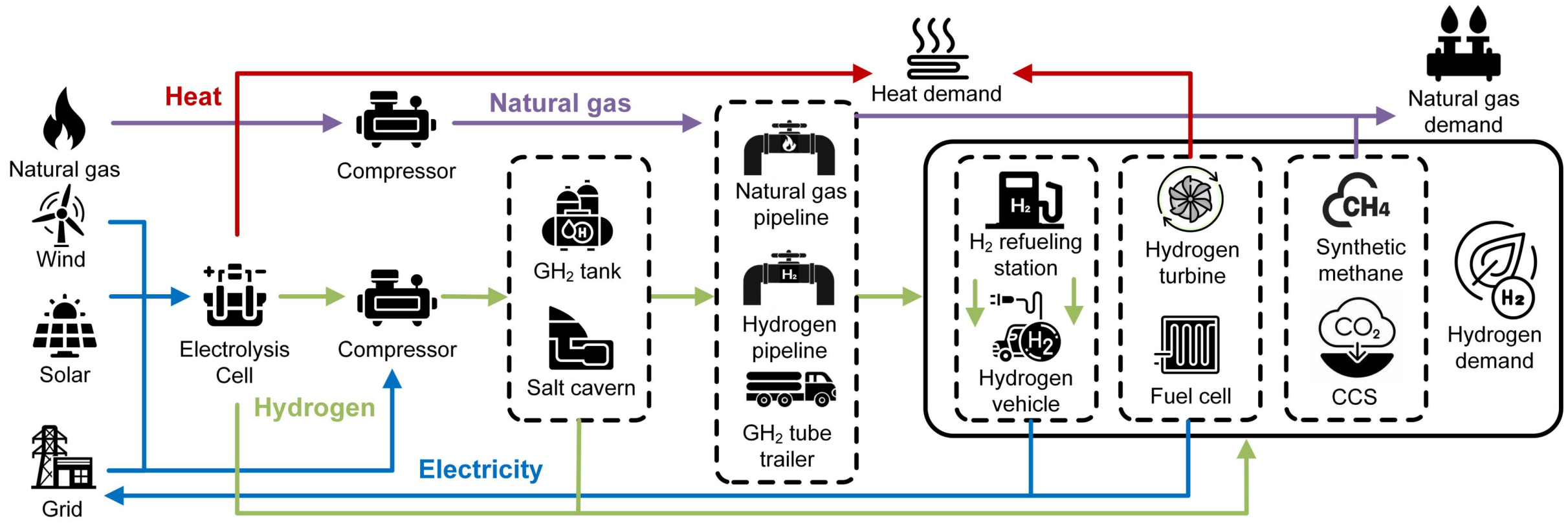

**Fig. 9.** Complete HEC

## 4. Modeling characteristics of HEC-MES planning

This section summarizes the main modeling features of HEC-MES, as shown in Table 2. To highlight cutting-edge research on collaborative planning between hydrogen and multiple other energy flows besides electricity, we have excluded literature that involved only electricity and hydrogen. We explore the sectors covered and network expansion content in HEC-MES. In addition, we review the spatio-temporal characteristics of the system and provide a detailed analysis of the sources and the modeling approach for uncertainty, as shown in Fig. 10.

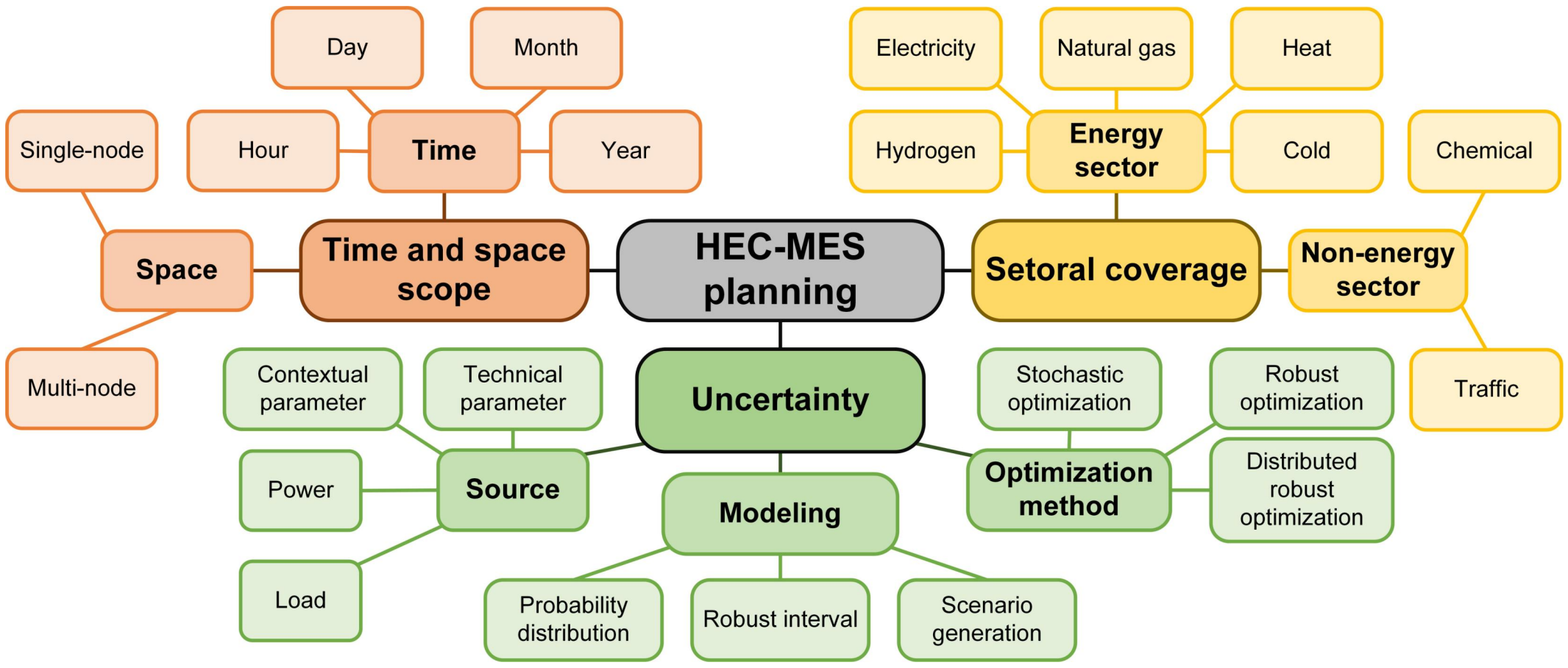

**Fig. 10.** The aspects related to the modeling characteristics of HEC-MES planning

## *4.1 Sectoral coverage*

HEC-MES involves a complex coupling of multiple sectors, which contributes to revealing the competitive relationship between multiple technologies [106]. Their synergistic optimization is a crucial technique for planning research, focusing on the interactive and energy conversion relationships between diverse sectors. It promotes the optimal allocation of multiple energy flows across time and space and realizes multi-energy complementarity and gradient utilization through rational planning of multiple energy-related equipment [27]. The research statistics of the sectors are shown in Fig. 11.

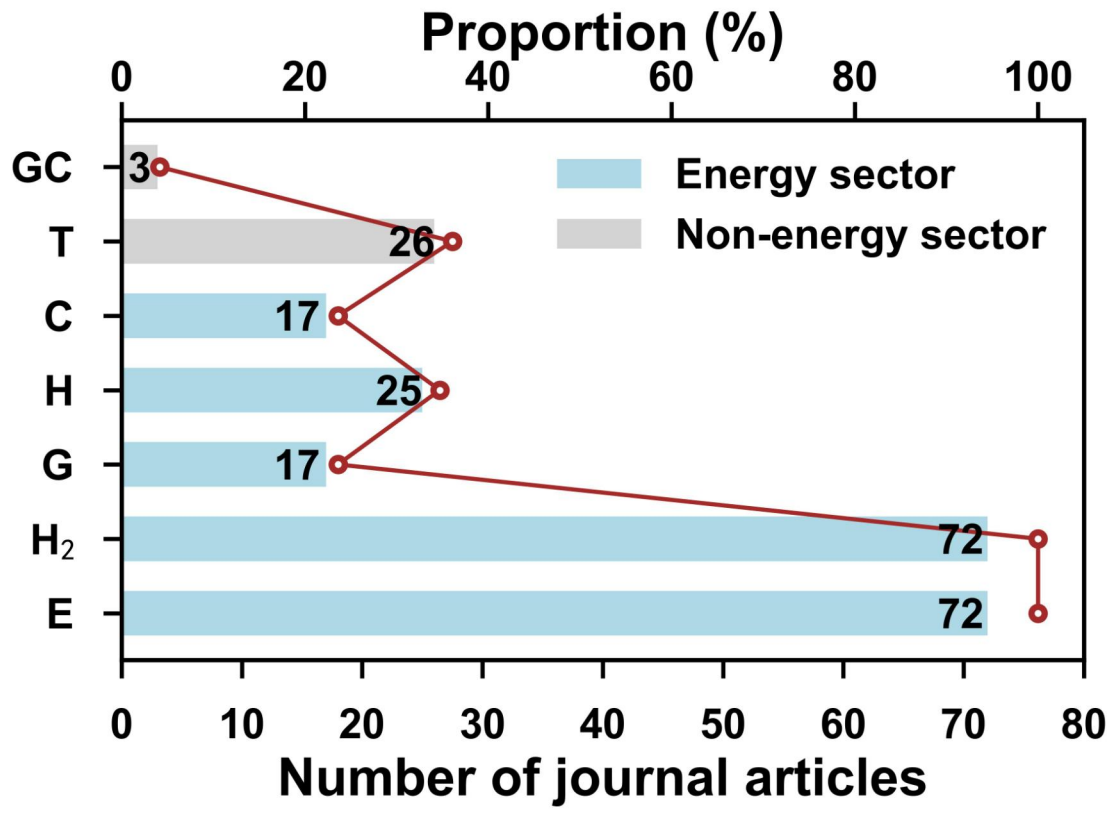

**Fig. 11.** Research statistics of the sectors

**Table 2 -** Modeling characteristics of HEC-MES planning

| Ref. | NE | Sector | | | | | | | Time | | Space | Uncertainty | | No. of objective function | | Form | Feature | Solution technology |
|---|---|---|---|---|---|---|---|---|---|---|---|---|---|---|---|---|---|---|
| | | E | G | $H_2$ | H | C | T | GC | Scale | RES | Scale | Source | Modeling | Single | Multi | | | |
| [24] | - | ✔ | ✔ | ✔ | ✔ | - | - | - | Y | Hr | Community | - | - | - | ✔ | MINLP | - | CPSO-GE |
| [57] | - | ✔ | ✔ | ✔ | ✔ | ✔ | - | - | TD (4,SEA) | Hr | MG | - | - | - | ✔ | LP | Bi-level | Gurobi+PSO |
| [59] | - | ✔ | ✔ | ✔ | ✔ | | - | - | Y+TD (18) | Hr | 4 zones in EUR | CP+TP | MC+PD | ✔ | ✔ | MILP | - | Matlab+Gurobi |
| [60] | - | ✔ | - | ✔ | ✔ | ✔ | ✔ | - | Y+TD (12) | Hr/D | Community | L+CP | EC+SG | ✔ | - | MILP | Bi-level | Matlab+Yalmip+ Gurobi+RT-GWO |
| [61] | - | ✔ | ✔ | ✔ | ✔ | ✔ | - | - | TD (12,SEA) | Hr/D | MG | P+L+N-1 | RO | ✔ | - | MILP | Two-stage | Nested-C&CG |
| [62] | - | ✔ | ✔ | ✔ | ✔ | - | - | - | Y+TD (3-72) | Hr | Zurich | - | - | - | ✔ | MILP | - | Cplex |
| [64] | - | ✔ | - | ✔ | ✔ | - | ✔ | - | Y+TW | Hr/M | MG | - | - | - | - | - | - | Simulation+Matlab+Simulink |
| [65] | - | ✔ | - | ✔ | ✔ | ✔ | ✔ | - | Y | Hr/D/M | 5 Nation | - | - | - | - | - | - | Simulation |
| [66] | ✔ | ✔ | - | ✔ | ✔ | ✔ | ✔ | - | TD | Hr/D | Fujian (9) | P+L | CCP+ DBSCAN | ✔ | - | MILP | Two-stage | Cplex |
| [67] | - | ✔ | - | ✔ | ✔ | ✔ | - | - | Y | Hr | Industrial park | - | - | ✔ | - | MILP | - | Matlab+Yalmip+ Cplex |
| [74] | - | ✔ | - | ✔ | ✔ | ✔ | - | - | Y | Hr/D/M | HRS | L | HOMER | ✔ | - | MIQCP | - | Matlab+Gurobi |
| [77] | ✔ | ✔ | ✔ | ✔ | ✔ | - | ✔ | - | Y | Hr | Germany (10) | - | - | - | ✔ | MILP | - | GAMS+Cplex |
| [78] | ✔ | ✔ | ✔ | ✔ | - | - | ✔ | - | TW (SEA) | Hr/D | MG (6/188-bus) | P+L | RO | ✔ | - | MILP+LP | - | Cplex+C&CG |
| [81] | - | ✔ | - | ✔ | ✔ | ✔ | - | - | TD (4,SEA) | Hr | EH-IES | - | - | - | ✔ | MIP | Bi-level | R&D+Cplex |
| [82] | - | ✔ | - | ✔ | ✔ | ✔ | - | - | Y | Hr | 12 city in China | - | - | ✔ | - | MILP | - | Python/GAMS+ Gurobi/Cplex/Mosek |
| [87] | - | ✔ | - | ✔ | ✔ | ✔ | - | - | TD (2000) | Hr | H-RE-CCHP system | CP+L | MC+DRO | - | ✔ | MILP | Two-stage | SDT+AUGMECON |
| [88] | ✔ | ✔ | ✔ | ✔ | - | - | - | ✔ | Y | Hr/Y | SA (43-node) | - | - | ✔ | - | LP | Dynamic | GAMS+Cplex |
| [91] | - | ✔ | - | ✔ | ✔ | - | ✔ | - | TD | Hr | Espoo | P+L+TP | SBSO | - | ✔ | MINLP | - | GWO-SCA |
| [93] | ✔ | ✔ | ✔ | ✔ | ✔ | - | - | - | TD (12) | Hr/M | Multi-MG (3) | P | ST | ✔ | - | MILP | Two-stage | FICO ®Xpress+NNB |
| [95] | - | ✔ | - | ✔ | ✔ | - | ✔ | - | Y | Hr/D/Y | Croatian | - | - | ✔ | - | LP | - | Python+Gurobi |
| [97] | - | ✔ | ✔ | ✔ | ✔ | ✔ | ✔ | - | TD (4,SEA) | Hr | IES site | P+L | GAN | - | ✔ | MILP | - | Gurobi |
| [101] | - | ✔ | ✔ | ✔ | ✔ | ✔ | - | - | Y | Hr | Community | - | - | - | ✔ | MINLP | - | MRM+PSO |

* *Corresponding author.*
E-mail address: tlu@sdu.edu.cn (T. Lu).

| | | | | | | | | | | | | | | | | | |
|---|---|---|---|---|---|---|---|---|---|---|---|---|---|---|---|---|---|
| [103] | - | ✔ | ✔ | ✔ | ✔ | ✔ | – | - | Y | Hr | Campus | - | - | ✔ | – | MILP | Three-stage | Python+Calliope+ slover |
| [104] | - | ✔ | ✔ | ✔ | ✔ | ✔ | – | - | Y | Hr/M | 2 buildings | - | - | ✔ | – | - | - | Simulation+TRNSYS+Genopt |
| [105] | - | ✔ | ✔ | ✔ | ✔ | ✔ | – | - | TD | Hr | Gansu | CP | GMM+PWE+LHS | - | ✔ | MIP+LP | Bi-level+Two-stage | Matlab+Gurobi+FA+MTGS+COPRAS |
| [109] | ✔ | ✔ | ✔ | ✔ | ✔ | ✔ | ✔ | ✔ | TD (3,SEA) | Y | Italy (3) | - | - | ✔ | – | MINLP | - | Simulation |
| [110] | ✔ | ✔ | ✔ | ✔ | – | – | – | - | TD (5) | Hr | MG (24-bus)+ GS (12-node) | - | - | ✔ | – | MILP | - | GAMS+Gurobi |
| [111] | ✔ | ✔ | ✔ | ✔ | ✔ | ✔ | – | - | TD (3) | Hr | MG (39-node)+ GS (20-node)+ TS (6-node) | P | MC+SBE+Cartesian product | - | ✔ | MILP | - | Multi-Objective PSO |

E, electric power; G, natural gas; H2, hydrogen; H, heat; C, cold; T, traffic; GC, green hydrogen chemistry; NE, network expansion; RES, resolution; Y, year; Hr, hourly; CPSO-GE, chaotic variational particle swarm algorithm based on greed and elite retention strategies; TD, typical day; EUR, Europe; LP, linear programming; CP, context parameter; TP, technology parameter; PD, probability distribution; L, load; EC, eigenvalue clustering; SG, scenario generation; RT-GWO, random trigonometric grey wolf optimization; P, power; C&CG, column-and-constraint generation algorithm; TW, typical week; CCP, chance constraint programming; DBSCAN, density based spatial clustering of applications with noise; MIQCP, mixed integer quadratic constrained programming; D, daily; EH-IES, electricity-hydrogen integrated energy system; MIP, mixed integer programming; R&D, reformulation and decomposition algorithm; GAMS, general algebraic modeling system; H-RE-CCHP, hydrogen-involved total renewable energy combined cooling, heating and power system; SDT, strong duality theory; AUGMECON, augmented ε-constraint method; SA, South Africa; SBSO, scenario based stochastic optimization; SCA, sine-cosine algorithm; ST, scenario tree; NNB, Newton's numerical barrier method; GAN, controllable generative adversarial network; MRM, maximum rectangle method, PWE, Parzen Window Estimation; LHS, Latin hypercube sampling; GMM, Gaussian mixture model; FA, firefly algorithm; MTGS, Mahalanobis-Taguchi Gram-Schmidt method; COPRAS, comprehensive proportional assessment;GS, natural gas system; TS, thermal system; Cartesian product; SBE, synchronous back-substitution elimination method

### *4.1.1 Energy sector*

The HEC-MESs integrate not only electric power and hydrogen but also the energy sector covering natural gas, heat, and cold. Modeling relies on the multi-energy conversion relationship mainly achieved through energy coupling devices. In HEC-MES, multi-energy coupling devices such as ECs, FCs, and HTs can all realize the coupled conversion of electricity, heat, and hydrogen [61,64]. In addition, HBs also allow for hydrogen-heat conversion [93]. Natural gas can reform into hydrogen. Hydrogen and $CO_2$ can be synthesized into methane with CCS to reduce carbon emissions [88]. Electric refrigeration equipment [67] and absorption chillers [82] enable electricity-cooling and heat-cooling conversion, respectively.

### *4.1.2 Non-energy sector*

HECs involve a wide range of technological and industrial sectors, so their introduction promotes the coupling of energy systems with non-energy sectors [137]. The traffic sector in HEC-MES focuses on electric vehicles and HVs. HVs can be categorized into hydrogen-fueled vehicles [107] and HFCVs [118]. Literature [91] proposed an improved smart charging strategy to simulate the charging process of electric vehicles from the grid, which can reduce the installed capacity of energy storage and regulate the balance between the supply and demand of electricity. Both electric and hydrogen vehicles serve as flexible loads and energy storage devices in HEC-MESs. Reference [65] considered that HFCVs can generate electricity by consuming hydrogen to supplement the power shortage of the grid at night and achieve power balance.

Furthermore, green hydrogen chemicals such as synthetic methane and ammonia have great potential in MES. Literature [88] proposed a hydrogen utilization mode of EC-synthesis methane-GT, which can effectively reduce carbon emissions by incorporating CCS technology. Research [138] utilized the wind-electricity-hydrogen-ammonia process to model ammonia demand as a flexible electrical load for grid co-planning. This approach effectively improves wind energy utilization and reduces grid expansion investment. Literature [139] proposed ammonia as a hydrogen carrier for large-scale, long-distance storage and transportation at a much lower cost than liquid hydrogen. However, few HEC-MES planning studies have involved green hydrogen chemistry. Collaborative planning of HEC-MES integrating ammonia remains a valuable research direction to be explored.

* *Corresponding author.*
E-mail address: tlu@sdu.edu.cn (T. Lu).

### 4.1.3 Network expansion of sectors

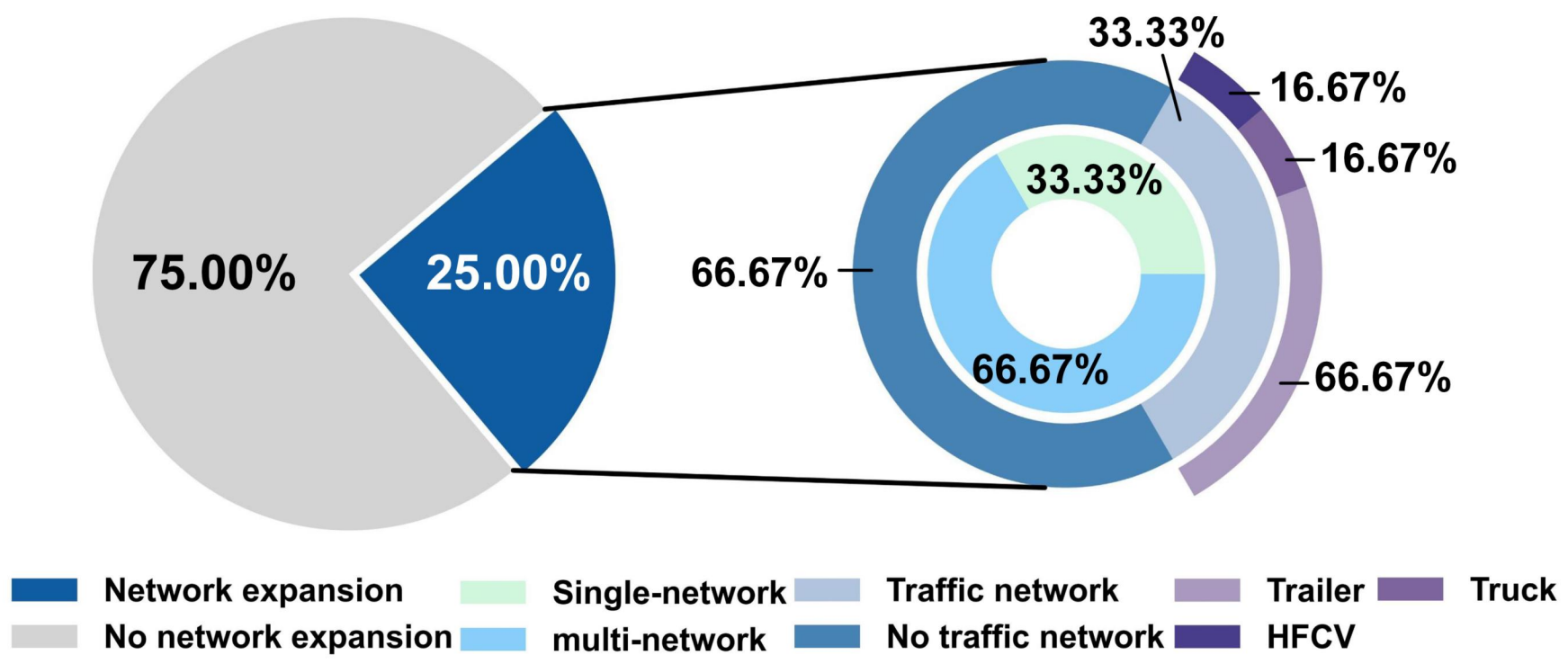

**Fig. 12.** Research statistics of network expansion

The research statistics of network expansion are shown in Fig. 12. Some literature suggests that collaborative planning of energy equipment and networks in HEC-MES can optimize the inter-regional energy transmission capacity and reduce the redundant installed energy equipment capacity [88]. It can also reduce investment costs, alleviate energy delivery congestion, and improve energy efficiency. Some researchers only consider a single hydrogen transportation method or even a single energy network expansion. For example, study [66] examined only the tube trailer expansion of a hydrogen transportation network. Other researchers have explored the synergistic expansion optimization of multi-energy networks, such as in literature [77] and [78], which expanded the electric-hydrogen network to promote the optimal allocation of heterogeneous energy in a complementary and cross-space manner.

Furthermore, the road transportation of hydrogen is dependent on traffic networks. Therefore, collaborative multi-network planning for electricity, hydrogen, and traffic is essential for coordinating the flow of electricity and hydrogen energy and traffic. Literature [79] coupled electricity, hydrogen, and transportation networks through ECs and hydrogen-fueled vehicles. It explored the synergies of multiple networks to optimize the flexibility and economy of the energy system. Yuanzheng Li et al. [84] established a multi-network planning structure considering the coupling constraints of power flow in an electric network, mass flow in a hydrogen network, and traffic flow in a traffic network to plan energy supply facilities. The results demonstrate that the structure effectively optimizes the cost, renewable energy utilization, voltage deviation, and transmission loss performance while ensuring the safety of the electric-hydrogen network.

As more sectors and energy network expansions are introduced, the complexity and solving difficulty of the model increase. All sector coupling and multi-network expansion have not been considered simultaneously in every HEC-MES planning study we reviewed. Addressing the challenge of reconciling the fineness of multi-sector coupling with the model's complexity is necessary for future HEC-MES planning studies.

### *4.2 Time and space scope*

Time and space are the basic features of planning models, limiting the scope of the study. They are closely related to the model's computational speed and problem-solving difficulty.

#### *4.2.1 Time scope*

Temporally, planning studies for HEC-MES that consider seasonal hydrogen storage typically use the year as the time scale, with a resolution of monthly, daily, and hourly intervals. Annual resolution is commonly used to study transition paths and guidelines for large-scale energy system transitions [109]. For example, literature [88] and [95] proposed a multi-year planning analysis with an annual resolution to analyze the inter-annual installed capacity changes during the energy system transition process. However, the magnitude and complexity of the planning problem increases with increasing time scale. Therefore, long-term planning with hourly resolution suffers from the difficult and time-consuming nature of solving this problem. The most common simplification is planning based on typical days derived from historical data [81,91]. HEC-MES planning is often divided into 4 typical days by season [57] or 12 typical days by month [61]. Because of their unique and discrete typical days, these methods fail to simulate medium- and long-term storage properties precisely. Paolo Gabrielli et al. [62] proposed two MILP models that distinguish different decision variable types. These simplify the integer decision variables for equipment based on typical days, while other decision variables maintain hourly accuracy throughout the year. The above methods are applied in the literature [59] and [60], which validate that they can effectively reduce model complexity and computation.

#### *4.2.2 Space scope*

Spatially, HEC-MES studies can range from the sizes of countries [77] and continents [88] to distribution grids [76], parks [67], or even hydrogen refueling stations [74]. Based on the presence or absence of a transmission network, we categorize HEC-MES into single-node systems without a network and interconnected multi-node systems. Most studies focus on the former without considering any transmission network expansion or energy interaction and ignore the geographic configuration and renewable energy endowment impacts of equipment planning. There is also some literature focusing on small-scale islanded energy systems with self-sufficiency, such as residential buildings [117], rural areas [121], and ships [140]. Reference [125] designed islanded hybrid renewable energy systems for residential buildings with ECs, FCs, and HSTs. Literature [102] proposed various topologies for PV-based microgrids for residential buildings and analyzed the optimal capacity configuration for both grid-connected and islanded states. The results show that the islanded mode is more expensive but improves reliability, autonomy, and energy efficiency and achieves zero-carbon emissions. Both the resource distribution [75] and load density [108] vary between regions. Researchers usually simplify a district into a single node to reduce complexity. For example, literature [77] and [109] divided Germany and Italy into multi-node systems by region, respectively. For multi-node systems, the degree of refinement of the transmission network model relates to the size of its spatial extent. Smaller low-voltage energy systems usually consider detailed flow models, which are generally categorized as either direct current flow [78] or alternating current flow [93]. The latter considers active and reactive power, making its planning results more in line with reality. Correspondingly, the problem becomes more complex and usually requires convexification and linearization operations.

### *4.2.3 General relationship between the planning model and different resolutions*

Temporal and spatial resolution contributes to the complexity and solution difficulty of the planning model and the accuracy of the results. Literature [88] modeled the level of spatial resolution by different numbers of nodes. Comparing the 30- and 43-node scenarios, there is less than 1% error in the planning results and a 30% reduction in the solution time. A reasonable selection of spatio-temporal resolution can improve both the accuracy of the results and the solution efficiency. However, few studies have focused on assessing the influence of spatial and temporal resolutions on planning.

## *4.3 Uncertainty*

With the enhancing coupling of multiple energy sources and high penetration of renewable energy access, complex and diverse uncertainties can significantly influence HEC-MES planning.

### *4.3.1 Uncertainty source*

The research statistics of uncertain sources are shown in Fig. 13. In the reviewed literature, the uncertainties can be categorized into four main types by source: contextual parameters, technical parameters, power, and load. Uncertainty in contextual parameters can be divided into policy context and environmental context. The former usually includes electricity and gas prices, feed-in tariffs, and carbon taxes [59]. On the other hand, the latter include irradiance, ambient temperature, and wind speed [60,87]. The uncertainty of meteorological data mentioned above can directly affect renewable energy output. The technical parameters include cost, lifetime, charging/discharging power, and energy loss [59,91]. The power uncertainty mainly arises from the volatility of renewable energy, such as wind, hydro, and solar [60,93]. This type of uncertainty requires a high degree of system flexibility and usually results in the redundant capacity of flexible resources such as ECs and HSs. Uncertainty exists for many load forms, such as electricity, heat, cold, and hydrogen, especially for flexible and adjustable loads such as HVs [61,97]. In addition, uncertainty due to device N-1 failures can affect the safety and reliability of HEC-MES [61]. However, considering N-1 failures in planning is a current research gap due to the variety of device types and the complexity of safety calibration calculations.

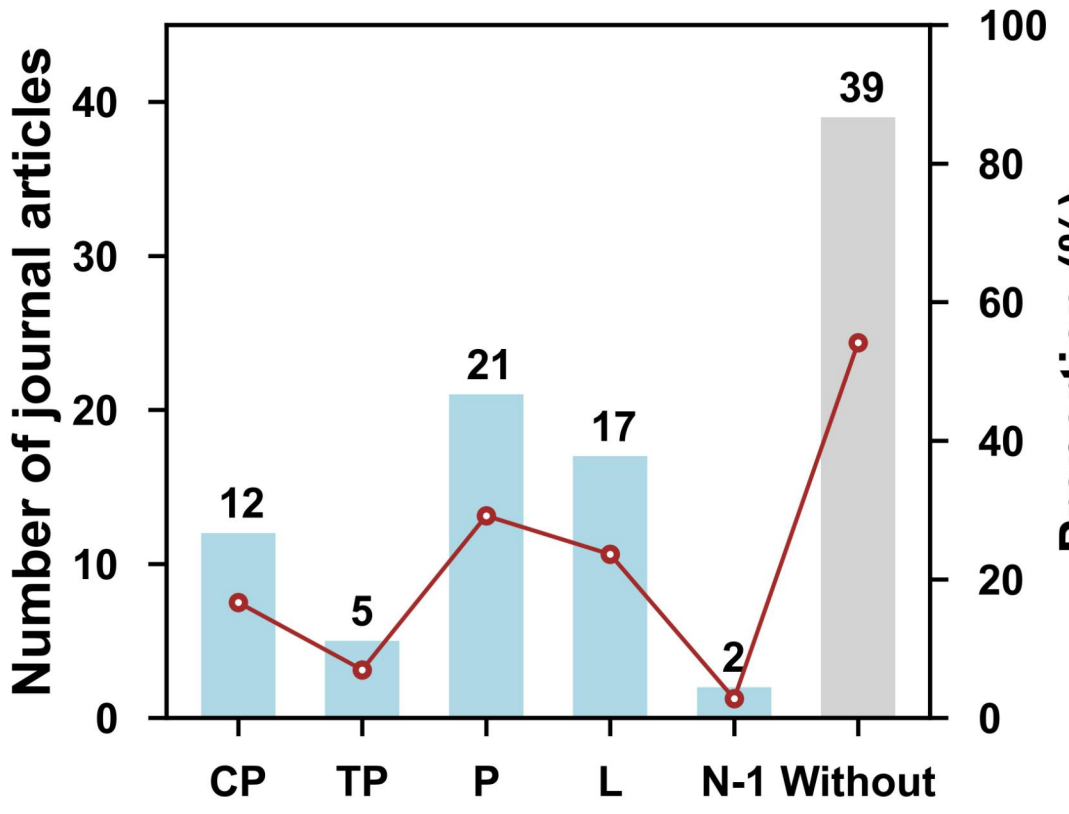

**Fig. 13.** Research statistics of uncertainty sources

### *4.3.2 Uncertainty modeling*

Modeling uncertainty is the basis for quantifying its impact on HEC-MES collaborative planning. The literature surveyed utilizes three main categories of uncertainty modeling methods: probability distribution methods [59], robust interval methods [60], and scenario generation methods [61].

The probability distribution method typically employs a probability function to characterize the range of values and the occurrence probability of values for the uncertainty of a parameter. Reference [59] utilized a uniform distribution to describe the uncertainty of electricity and gas prices and a program evaluation and review technique distribution to describe the uncertainty of the discount rate. Literature [74] utilized HOMER software to set stochastic volatility for hydrogen demand uncertainty.

The robust interval method utilizes the uncertainty set, which is a bounded set, to bound the fluctuation range of parameter uncertainty [78]. Research [60] used the upper and lower deviations of forecasts to define the uncertainty set for describing the uncertainty in renewable energy output and electricity, heat, cooling, and hydrogen load.

The scenario generation method describes uncertainty through discrete scenarios generated from historical data or random probability sampling. Literature [61] performed eigenvalue clustering on a time series of uncertain parameters such as irradiance, wind speed, load, and scenario reduction and generation of clustering results. Reference [87] constructed fuzzy sets considering irradiance and multi-energy load uncertainties and then generated scenarios with normally distributed parameter prediction errors using the MC method. For large-scale and complex data, scenario generation methods can also utilize artificial intelligence techniques such as deep learning to improve accuracy and computational efficiency [141]. For instance, literature [97] utilized improved controllable generative adversarial network techniques to characterize uncertainties such as renewable energy outputs and loads and then efficiently generated both conventional and extreme operating scenarios.

### *4.3.3 Uncertainty optimization method*

Methods for planning optimization after considering uncertainty modeling can be classified into three main categories: stochastic optimization (SO) [59], robust optimization (RO) [61], and DRO [87]. SO is typically based on plentiful random scenarios from historical data and probability distributions. Although the results are highly accurate, large-scale variables and constraints can lead to solving difficulties. Compared to SO, RO is computationally simple but suffers from overly conservative results. DRO combines the advantages of both methods by incorporating historical data to construct an uncertainty probability distribution. It facilitates computational solutions while ensuring the accuracy of the results. However, this method is currently less applied in HEC-MES planning with uncertainty, which requires further research.

## 5. Methodology of HEC-MES planning

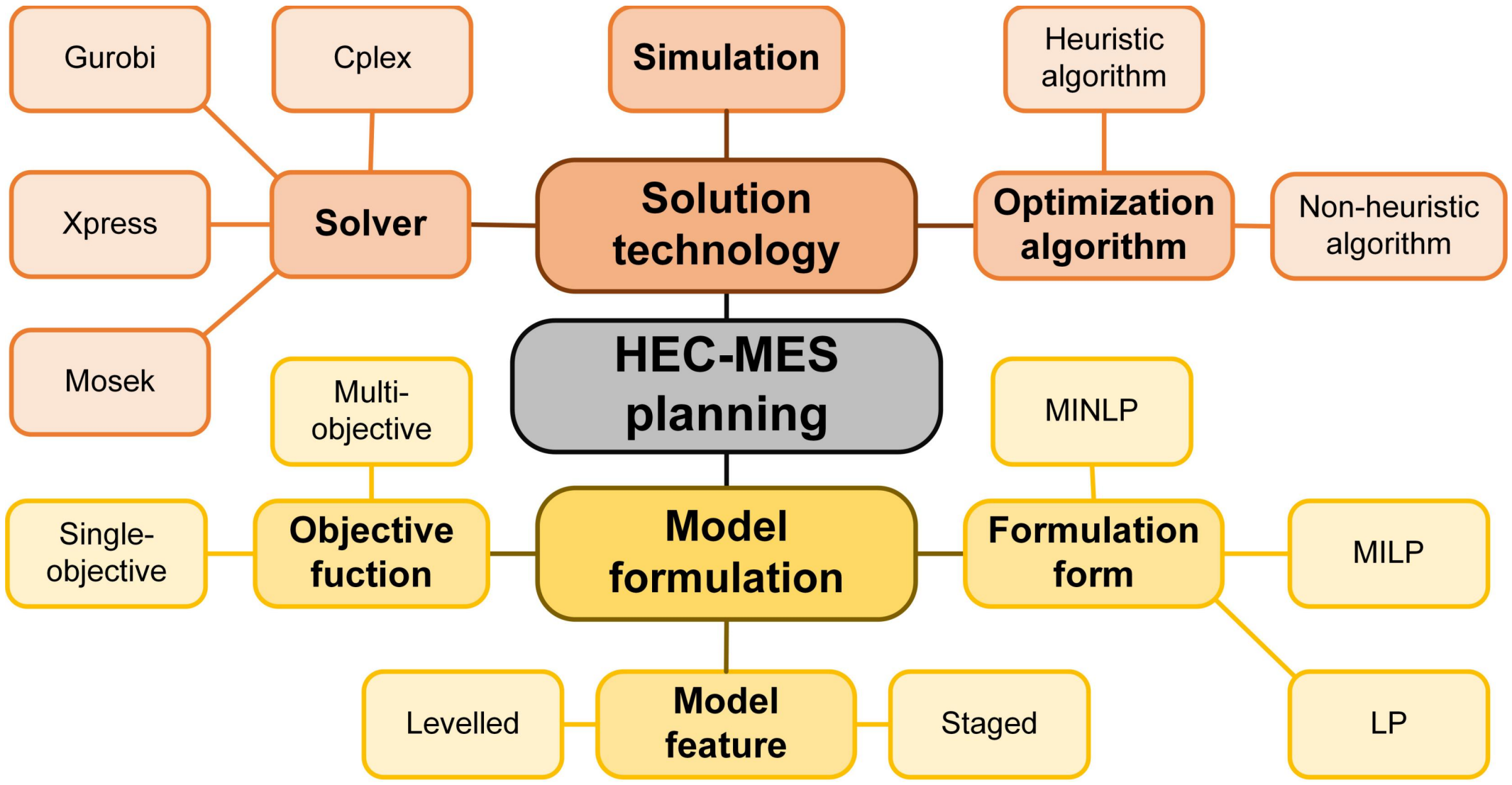

**Fig. 14.** The aspects related to the methodology of HEC-MES planning

This section summarizes the main HEC-MES planning methodology, including a detailed analysis of the model formulation and solution technologies, as presented in Fig. 14.

## 5.1 Model formulation

The model formulation is the basis of the HEC-MES planning solution, thus we mainly analyze the objective function, the formulation form, and the model features.

### 5.1.1 Objective function

Similar to other energy systems, in the literature reviewed, the most common form of the objective function for HEC-MES planning is the single-objective of minimizing economic indicators. The economic indicator is usually the total annual net present value cost (TANPC), which typically includes investment costs and O&M costs considering the discount rate [66]. Some literature has also incorporated transaction costs with the grid, hydrogen sales gains, and penalty costs [61]. Penalty costs are commonly related to energy abandonment or low-carbon policies for reducing energy curtailment rates and carbon emissions [67]. In addition, some studies have utilized the life cycle cost as a planning objective, considering the cost of retirement and replacing hydrogen energy equipment to quantify economics more comprehensively [74].

Several studies reviewed have performed multi-objective optimization to achieve the optimal comprehensive multi-dimensional benefits of HEC-MES planning. Generally, these studies considered two or more performance metrics covering economic, environmental, reliability, and energy efficiency [57,97]. Study [91] combined TANPC minimization and levelling cost minimization for dual objective planning. Literature [24] synthesized economic, environmental, and energy efficiency indicators for three-objective planning. Furthermore, Pareto analysis is used to obtain the Pareto frontier, which is combined with the Technique for Order Preference by Similarity to the Ideal Solution method to set weights for each indicator for selecting the optimal planning solution. The suitable selection of metrics is crucial for multi-objective planning. Reference [101] characterized energy efficiency by the primary energy saving rate and reliability using the grid dependence reduction rate. On the other hand, literature [120] selected the probability of power supply loss to characterize reliability.

### 5.1.2 Formulation form

As mentioned previously, HEC-MES involves tightly coupled sectors, multiple energy devices, and especially hydrogen energy devices with complex dynamic characteristics. For instance, ECs have non-linear efficiency profiles [74]. Moreover, the number of devices [91] and operational state changes inevitably introduce integer variables, such as the start/stop states of ECs [67] and charging/discharging states of HSTs [60]. The planning problem for HEC-MES is characterized by high dimensionality, non-linearity, and non-convexity. Its formulation is a mixed integer non-linear programming (MINLP) model with high accuracy but is computationally intensive and difficult to solve. Therefore, many studies simplify it to a MILP model using convexification and linearization. For example, literature [61] and [93] utilized the planar approximate linear method for linearization. As shown in Table 2, the MILP model is the most common form in the literature reviewed, which balances accuracy and solution efficiency.

However, in large-scale and long-term planning scenarios, vast discrete integer variables and constraints can cause the MILP model to fail in direct solution. Relaxation of integer variables can simplify the problem into a linear programming (LP) model that is easier to solve. Research [78] adopted the relaxed linear continuous model of hydrogen energy trucks to simplify the calculation. However, it is essential to note that this process inevitably introduces errors, which may reduce accuracy.

### *5.1.3 Model features*

HEC-MES planning models are usually decomposed into levelled or staged structures for hierarchical optimization to reduce complexity and calculations.

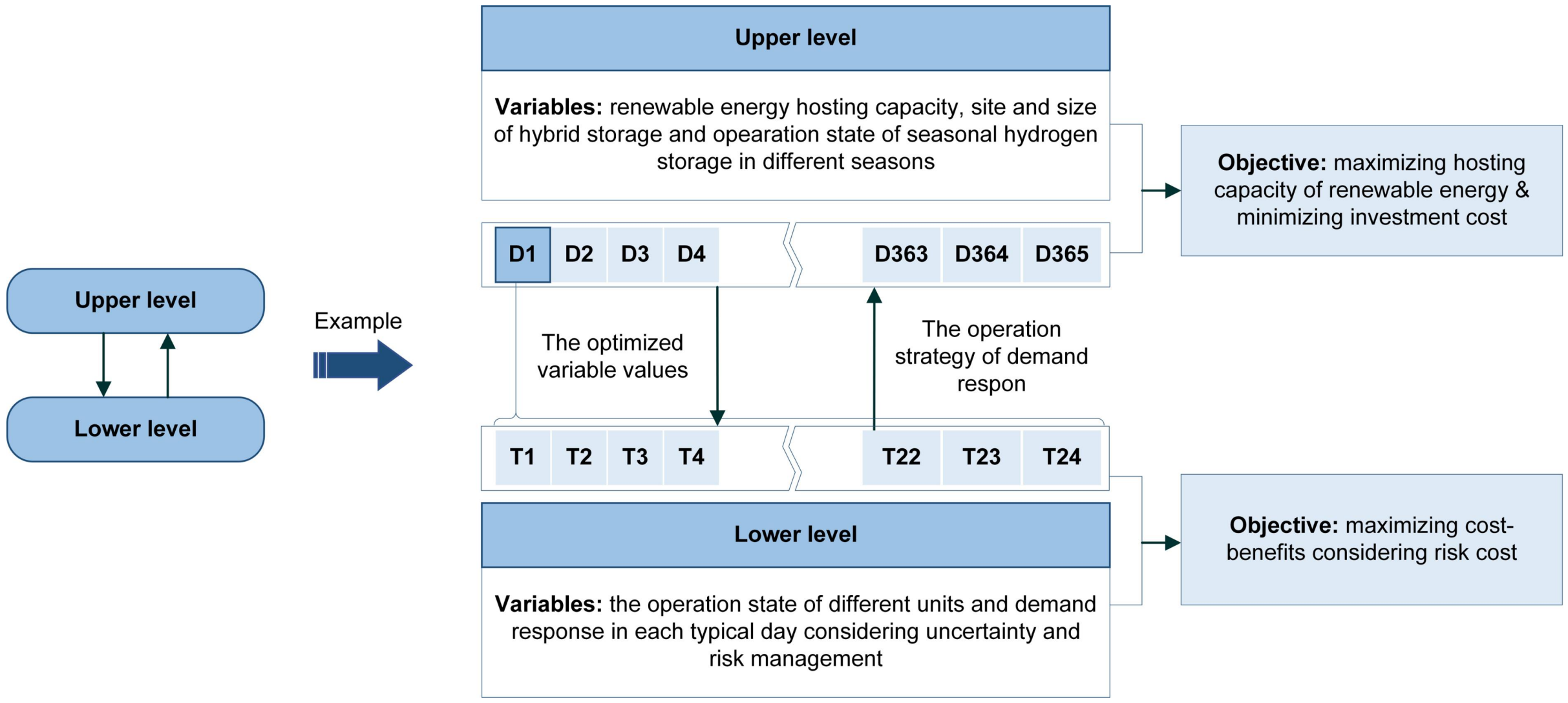

**Fig. 15.** Typical bi-level model structure

For levelling, every level interacts with each other to obtain optimal planning results. A bi-level model is a typical form in HEC-MES, as shown in Fig. 15 which illustrates the model framework of the literature [115]. In study [60], lower-level optimized scheduling minimizes daily operating and retirement costs and upper-level optimized capacity-setting results in maximizing life cycle costs. In the literature [81], the results of investment decisions in the upper level affected the LCOH objective of the lower level, and the operational output power of the lower level affected the TANPC objective of the upper level. In literature [107], the upper level performs a vehicle-to-grid scheduling optimization and inputs the results into the lower level model for solving the optimal capacity allocation of the microgrid.

For staging, there is an explicit sequence between stages. A two-stage model is a typical form of planning HEC-MES, which is usually used for RO or SO with uncertainty. Figure. 16 illustrates the two-stage model framework with an example from the literature [118]. The model determines the system structure and equipment capacity in the first stage and optimizes the charging/discharging power of the batteries and HFCVs after the uncertainty is revealed in the second stage. Similarly, reference [66] performed optimization of equipment siting and capacity in the first stage. After confirming parameter uncertainty, the second stage performed optimization of equipment output and inter-regional hydrogen transfer. In literature [61], the first stage made a deterministic investment decision, and the second stage optimized the equipment output under the worst-case scenario of uncertainty.

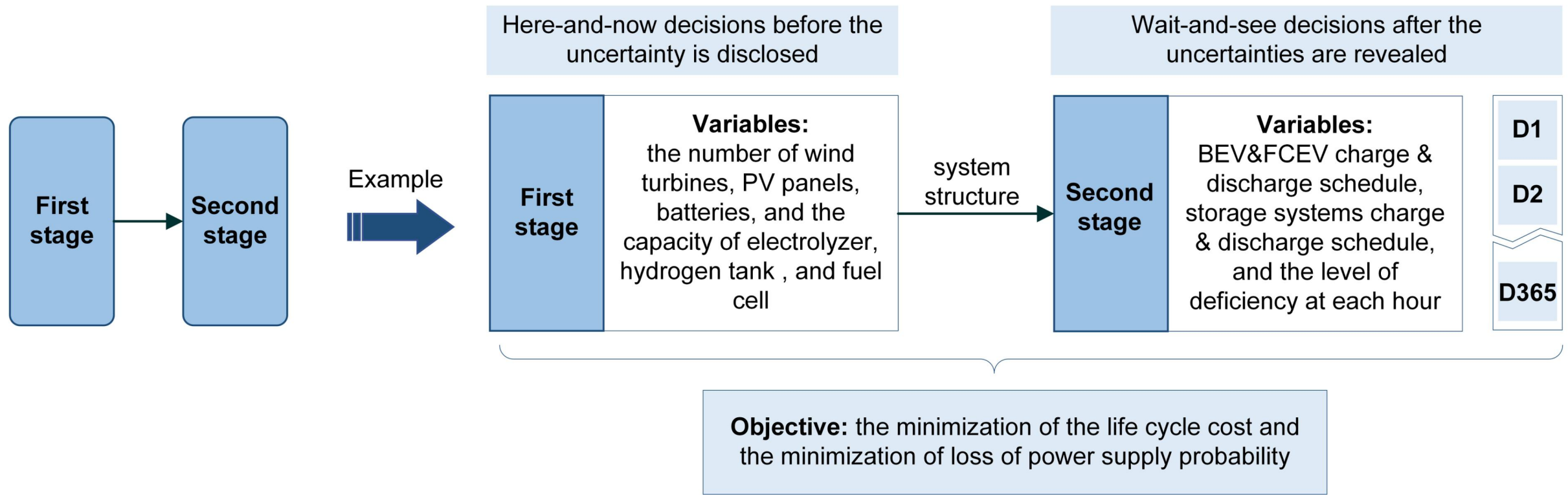

**Fig. 16.** Typical two-stage model structure

Levelling and staging are not contradictory, and a reasonable combination of both can form a better solution strategy. Literature [105] proposed an optimization framework with a bi-level procedure and two-stage decision-making, which can coordinate the conflict between multiple objectives and simplify the solution complexity

## *5.2 Solution technology*

As shown in Table 2, the solution technologies for HEC-MES planning can be divided into three main categories: optimization algorithms, solvers, and simulations. The research statistics of the solution technology are shown in Fig. 17.

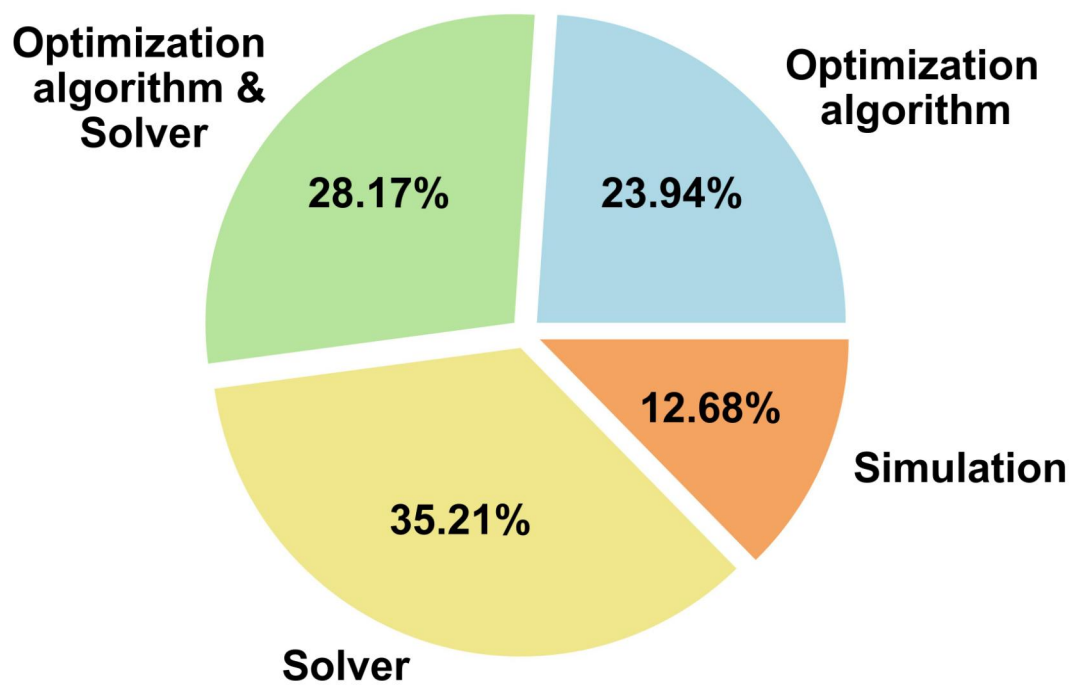

**Fig. 17.** The research statistics of the solution technology

### *5.2.1 Optimization algorithm*

With the solution efficiency and computational feasibility, optimization algorithms are preferred for solving large-scale MILP and complex MINLP problems. Heuristic algorithms, such as particle swarm optimization (PSO), greedy algorithm, and elite retention strategy (ERS), are a common category of optimization algorithms. Research [57] used the PSO algorithm to optimize the capacity of a hydrogen storage system. Non-heuristic algorithms, such as strong duality theory (SDT), ε-constraint algorithms, and Newton's numerical barrier method (NNB), belong to another category of optimization algorithms. Reference [93] used NNB to solve the planning results for analyzing the differences in the total cost and capacity of systems with different decarbonization pathways.

In addition to directly applying a single algorithm, some researchers have improved heuristic/non-heuristic algorithms. Literature [122] proposed an improved PSO algorithm that optimizes the accuracy of the optimal solution through the velocity equation and introduces a variable constriction factor to improve the convergence speed. Others combined multiple algorithms to achieve efficient solutions for complex models. Study [24] proposed a combinatorial algorithm named the chaotic particle swarm algorithm based on greedy and elite retention strategies for the multi-objective MINLP model. It combines the chaotic variational PSO, greedy variational strategy, and ERS to improve the optimality of finding efficiency and avoid falling into a local optimum. Literature [87] utilized SDT and augmented ε-constraint algorithm to identify the Pareto optimal solution set. The final solution was selected based on multi-objective weights with a fuzzy clustering algorithm. Reference [91] combined grey wolf optimization and the sine-cosine algorithm to solve the proposed MINLP model of an independent hybrid energy system. It improves the convergence efficiency and saves computation time compared to traditional algorithms.

### 5.2.2 *Solver*

Commercial solvers can efficiently find exact global optimal solutions of simple models, such as LP, MILP, and mixed integer quadratic programming (MIQCP) models. As shown in Table 2, commercial solvers widely used in HEC-MES planning include Cplex [62], Gurobi [97], Xpress [93], and Mosek [82]. They are usually combined with modeling software such as MATLAB [60], Python [95], and GAMS [88]. For complex and large-scale MILP problems, reasonable MIP-gap values set in commercial solvers can reduce the computational time and obtain sub-optimal solutions within an acceptable error range [59].

### 5.2.3 *Simulation*

Several researchers have employed simulations to analyze the planning of HEC-MES. Simulation is generally based on modules constructed by mathematical formulas to build MESs for planning analysis and commonly used simulation software such as MATLAB/Simulink [99] and HOMER[123]. Research [104] built a solar energy system integrating HST and FC in TRNSYS with Genopt solving to obtain the system capacity configuration when the life cycle cost is optimal. Some of the simulations disregard the optimization of the objective function and only set up a planning principle for capacity allocation. For example, literature [64] proposed capacity allocation principles and operational strategies that consider the spatial and temporal distribution of renewable energy and verified their feasibility by MATLAB/Simulink for modeling simulations. As another example, study [65] calculated the capacity of MESs based on simulation schemes and adjusted the planned capacity by whether or not the annual hourly energy flow balance was satisfied in several iterations.

## 6. Conclusion and future directions

With the further increase in renewable energy penetration, large-scale clean hydrogen production using WTs and PVs becomes possible and contributes to hydrogen energy development in the energy system. As a medium converted with multiple energy sources, HEC can realize the optimal allocation of heterogeneous energy sources across time and space. It can also boost the deep decarbonization of the energy system. However, HEC involves numerous devices and technologies, so there is an urgent requirement for rational and efficient planning methods to realize the deep integration of HEC with MESs. By focusing on the coupling of hydrogen energy with the energy sector, this paper provides a comprehensive overview of the planning research of HEC-MES. On the one hand, we introduce the concept of the HEC with each link and outline the status of planning for key hydrogen energy technologies in energy systems in recent years. On the other hand, we also systematically review the modeling features and methods of the HEC-MES planning problem, analyzing the involved sectors, spatial and temporal scopes, uncertainties, model formulations, and solution methods in depth.

Through a review of related literature, we summarize the gaps in the current research on HEC-MES planning and the directions that deserve to be explored in depth in future research, mainly as follows:

(1) Current research focuses on single or several hydrogen energy links, and few studies consider the complete HEC with all links, limiting the deep integration of hydrogen energy with MESs. In HEC-MES, the interactions between the various links of the HEC and their impact on planning are worth exploring;

(2) There are various types of hydrogen energy equipment and technologies, and most of the studies only consider a single equipment and technology for a particular link in the HEC while planning. The portfolio selection of HEC equipment that combines technical efficiency, flexibility, and economy still requires further research;

(3) Current studies have verified the advantages of multi-sectoral collaboration. However, the mechanism of cross-sectoral collaboration still needs to be investigated, and no study has yet comprehensively considered multi-energy coupling, multi-network, and multiple energy storage in a well-rounded manner. The effect of non-energy sectors such as HFCVs and green hydrogen chemicals in energy system planning is also an important direction for future research;

(4) Most current studies consider only one or two types of uncertainty and ignore the impact of N-1 fault uncertainty on planning security. Modeling methods incorporating information technologies such as deep learning and distributed robust optimization remain to be thoroughly investigated;

(5) A refined HEC model is established to accurately describe the synergies between various devices in each link to rationally and efficiently optimize the capacity allocation. An urgent challenge for future research is to find suitable optimization algorithms that effectively address the contradiction between model refinement and solution difficulty.

This review focuses on the synergistic planning study of MESs containing hydrogen energy chain, aiming to provide a reference for the promotion of top-down hydrogen energy chain applications and for exploring the shape of future low-carbon energy systems.

## Acknowledgements

This work was supported by the National Natural Science Foundation of China-Enterprise Innovation and Development Joint Fund (U22B20102), the Harvard Global Institute and Energy Foundation China.